\documentclass{article}
\pdfoutput=1

\usepackage{microtype}
\usepackage{graphicx}
\usepackage{subcaption}
\usepackage{booktabs} % for professional tables
\usepackage{multirow}    % For multirow cells
\usepackage{subcaption}  % For subtable environment
\usepackage{multicol}
\usepackage{graphicx}  % For adjusting table width if needed
\usepackage{array}
\usepackage{hyperref}
\usepackage{tabularx, booktabs}
\usepackage{adjustbox}

\usepackage[accepted]{icml2024}

\usepackage{amsmath}
\usepackage{amssymb}
\usepackage{mathtools}
\usepackage{amsthm}
\usepackage{footnote}
\usepackage[capitalize,noabbrev]{cleveref}

\theoremstyle{plain}

\theoremstyle{definition}

\theoremstyle{remark}

\usepackage[textsize=tiny]{todonotes}

\icmltitlerunning{Character-level Tokenizations are Powerful Priors for RNA Foundation Models}

\begin{document}

\twocolumn[
\icmltitle{Character-level Tokenizations as Powerful Inductive Biases for RNA Foundational Models}



\icmlsetsymbol{equal}{*}

\begin{icmlauthorlist}
\icmlauthor{Adrián Morales-Pastor}{nbd}
\icmlauthor{Raquel Vázquez-Reza}{nbd}
\icmlauthor{Miłosz Wieczór}{irb}
\icmlauthor{Clàudia Valverde}{nbd}
\icmlauthor{Manel Gil-Sorribes}{nbd}
\icmlauthor{Bertran Miquel-Oliver}{bsc}
\icmlauthor{Álvaro Ciudad}{nbd}
\icmlauthor{Alexis Molina}{nbd}

\end{icmlauthorlist}

\icmlaffiliation{nbd}{Department of Artificial Intelligence, Nostrum Biodiscovery, Barcelona, Spain}
\icmlaffiliation{irb}{Institute for Research in Biomedicine (IRB Barcelona), The Barcelona Institute of Science and Technology, Barcelona, Spain}
\icmlaffiliation{bsc}{Barcelona Supercomputing Center, Barcelona, Spain}
\icmlcorrespondingauthor{Alexis Molina}{alexis.molina@nostrumbiodiscovery.com}

\icmlkeywords{RNA language models, ChaRNABERT}

\vskip 0.3in
]



\printAffiliationsAndNotice{} % otherwise use the standard text.
%\printAffiliationsAndNotice{} % otherwise use the standard text.

\begin{abstract}
RNA is a vital biomolecule with numerous roles and functions within cells, and interest in targeting it for therapeutic purposes has grown significantly in recent years. However, fully understanding and predicting RNA behavior, particularly for applications in drug discovery, remains a challenge due to the complexity of RNA structures and interactions. 
While foundational models in biology have demonstrated success in modeling several biomolecules, especially proteins, achieving similar breakthroughs for RNA has proven more difficult. Current RNA models have yet to match the performance observed in the protein domain, leaving an important gap in computational biology.
In this work, we present ChaRNABERT, a suite of sample and parameter-efficient RNA foundational models, that through a learnable tokenization process, are able to reach state-of-the-art performance on several tasks in established benchmarks. We extend its testing in relevant downstream tasks such as RNA-protein and aptamer-protein interaction prediction.
Weights and inference code for ChaRNABERT-8M will be provided for academic
research use. %at: https://github.com/NBDsoftware/ChaRNABERT. 
The other models will be available upon request.
%including, but not limited to, prediction of RNA-protein and aptamer-protein interactions, RNA sequence generation and folding.

\end{abstract}

\section{Introduction}

RNA is known to be a pivotal molecule in molecular biology and research on its functions has significantly transformed our understanding of gene expression, regulation, and therapeutic potential. Unlike DNA, which primarily serves as a stable repository of genetic information, RNA is dynamic and versatile, taking on multiple roles within cells. It acts not only as the messenger that conveys genetic instructions from DNA to the protein synthesis machinery (mRNA) but also as a regulator of gene expression through molecules like microRNAs (miRNAs) and small interfering RNAs (siRNAs) \citep{micrornareg}. Additionally, some RNAs function as catalysts in biochemical reactions, exemplified by ribozymes \citep{ribozymesreview}. These unique structural and functional properties enable RNA to influence nearly every aspect of cellular biology, making it essential to both fundamental life processes and modern medical applications.

Recent breakthroughs in RNA biology have opened new avenues for therapeutic interventions, turning RNA into a powerful tool for combating diseases previously considered undruggable by conventional treatments. RNA therapeutics, such as RNA interference (RNAi) for gene silencing and mRNA vaccines that instruct the immune system, offer immense potential. The rapid development and deployment of mRNA vaccines during the COVID-19 pandemic showcased the speed, flexibility, and effectiveness of RNA technology in addressing global health crises \citep{vaccinesreview}. Beyond vaccines, RNA-based therapies are being explored to treat a wide range of conditions, including genetic disorders, cancer, neurodegenerative diseases, and metabolic syndromes. Techniques like antisense oligonucleotides (ASOs), siRNAs, and RNA aptamers leverage RNA molecules to modulate gene expression, inhibit harmful proteins, or even repair genetic mutations \citep{Rnatherapeutics}. Furthermore, RNA-guided genome editing tools, such as CRISPR-Cas systems, represent a new frontier in precision medicine by targeting RNA directly and offering unparalleled specificity in gene modulation without permanently altering DNA \citep{crisprcasreview}.

However, despite these advancements, challenges remain particularly in enhancing RNA stability, improving delivery mechanisms, and reducing off-target effects \citep{rnachallenges}. As RNA-mediated therapies advance, there is an increasing need for methodologies that not only rationalize RNA behavior and properties but also scale effectively to meet the demands of the field. Computational biology plays a critical role in predicting RNA behavior prior to experimental validation. Several software tools have been developed to predict RNA features, with ViennaRNA \citep{viennaRNA} being one of the most prominent. While the primary focus of RNA tool development has been on predicting properties, molecular simulations have played a crucial role in studying RNA’s dynamic behavior and validating its secondary and tertiary structures \citep{mdrna}. However, these simulations are often resource-intensive, requiring significant engineering expertise and deep knowledge of the modeled systems. Moreover, they may not always provide the resolution needed to capture the full complexity of RNA behavior. 

In contrast, the emergence of artificial intelligence in RNA research holds the promise of expanding the field, offering appealing opportunities that surpass traditional computational methods. AI models, especially at the inference stage, are dramatically more efficient to run and can tackle a vast array of tasks, many of which can be learned directly from RNA sequence data alone. While protein language models (pLMs) have revolutionized our understanding of protein folding, function, and design with models like Evolutionary Scale Modeling (ESM) \citep{esm2} excelling in applications ranging from drug discovery to protein engineering, AI applications for RNA have developed more gradually. Protein models have rapidly expanded their use cases, whereas RNA-focused models have often specialized in narrower tasks, such as predicting RNA splicing patterns or optimizing codon usage, limiting their broader impact. 

Recent efforts to create general-purpose RNA models show promise but still lag behind the advances seen in protein modeling. Even the most advanced RNA models, which have achieved state-of-the-art performance in areas like secondary structure prediction \citep{rinalmo}, have yet to achieve the transformative success that AI has brought to protein science.

This work aims to take a step toward establishing a robust RNA foundation model, ChaRNABERT, capable of performing a wide range of downstream tasks. Our approach is built on two core concepts: first, utilizing learnable tokenization to move beyond human-curated motif-selection, which may not be optimal for a biomolecule like RNA; and second, training on a diverse set of RNA types to ensure the model's ability to generalize across different types and tasks. 
To assess the ChaRNABERT's performance in relevant scenarios, we evaluate its performance on an standardized benchmark, expand the benchmark with additional tasks, and test its capabilities in high-impact applications such as aptamer interaction prediction. % RNA-protein interaction, inverse folding and RNA tertiary structure prediction.% and RNA sequence generation.

\section{Related Work}

The rapid advancement of RNA language models has tried to mirror the transformative impact of language models in protein science, aiming to decode the "language" of RNA sequences. These models endeavor to capture the underlying patterns, structural motifs, and functional elements inherent in RNA, thereby facilitating breakthroughs in structure prediction, functional annotation, and therapeutic design. The progression of these models reflects a concerted effort to overcome the unique challenges posed by RNA's structural diversity and functional versatility.

Early pioneers in this field, such as RNA-FM \citep{rnafm} and RNABERT \citep{rnabert}, laid the foundational groundwork for RNA language modeling. RNA-FM was one of the first general-purpose models designed for non-coding RNA (ncRNA) sequences. RNA-FM, a 100M parameter model, was trained on a previous RNAcentral release, encompassing 23 million samples. This model demonstrated the potential of language models to learn directly from RNA sequences, enabling tasks like secondary structure prediction and functional annotation. Concurrently, RNABERT emerged with a focus on structural alignment and clustering of ncRNA. By incorporating partial multiple sequence alignments from RNAcentral and the Rfam 14.3 dataset, totaling over 762 thousand sequences, RNABERT leveraged evolutionary information to enhance its ability to discern structural similarities among RNA molecules. This integration of evolutionary data marked a step toward understanding RNA structure-function relationships with a special focus over clustering and alignment.

Building on these foundations, models like UNI-RNA \citep{unirna} sought to scale up both in model complexity and dataset size. UNI-RNA featured 400M parameters and was trained on an expansive dataset of 1 billion sequences from RNAcentral, the Nucleotide Collection (nt), and Genome Warehouse (GWH). Aiming to be a universal RNA model, UNI-RNA endeavored to capture a broad spectrum of RNA types and functions, enabling the modeling of very long RNA sequences without truncation.

Application specific models also made significant contributions. RNA-MSM \citep{rnamsm} introduced a novel approach by directly utilizing evolutionary information from multiple sequence alignments to model ncRNA sequences, benchmarking a diverse array of dowstream tasks. SpliceBERT \citep{splicebert} addressed the critical aspect of RNA splicing in precursor messenger RNA (pre-mRNA), aiding in the prediction of splice sites and alternative splicing events. These advancements underscored the importance of specialized models in tackling specific biological questions.

Models like CodonBERT, UTR-LM, and 3UTRBERT \citep{codonbert,5utrmodel,3utrbert} focused on different regions of mRNA, capturing codon usage patterns and post-transcriptional regulation mechanisms mediated by untranslated regions (UTRs). CodonBERT, concentrated exclusively on the coding sequences (CDS) of mRNA, employing codon-level tokenization to capture patterns crucial for gene expression optimization. UTR-LM and 3UTRBERT specialized in the 5' and 3' UTRs, respectively, enhancing our understanding of mRNA expression, translational efficiency, and gene regulation mediated by UTRs.

BigRNA \citep{bigrna} diverged from sequence-based models by integrating genomic context and utilizing thousands of genome-matched datasets. This approach underscored the importance of multi-omics data in capturing the complexity of RNA regulation in different cellular contexts, moving beyond sequence information to include expression patterns and regulatory interactions.

Despite these advancements, a noticeable gap remained when compared to the transformative impact of language models in protein science. Many existing RNA models were specialized or limited in scope, hindering their generalizability and broader applicability. Addressing this challenge, RiNALMo (RNA Integrated Language Model Optimization) \citep{rinalmo} emerged as a notable milestone in RNA language modeling. RiNALMo was designed to bridge this gap by providing a comprehensive and versatile framework capable of capturing the full complexity of RNA sequences and structures. This model employed a deep transformer-based architecture with attention mechanisms tailored specifically for RNA.

One of the key introductions of RiNALMo was its pre-training strategy. The model was trained on an extensive and diverse dataset that included a wide array of RNA sequences from databases such as RNAcentral, as well as experimentally derived structural data. This multimodal training approach allowed RiNALMo to learn rich representations that encapsulate both linear sequence information and three-dimensional conformations of RNA, bridging the gap between sequence and structure. Moreover, RiNALMo introduced a modification in comparison to standard tokenization methods that went beyond simple nucleotide or codon representations. By utilizing k-mer embeddings and incorporating secondary structure annotations, the model could understand folding patterns and motifs crucial for RNA function. 

In practical applications, RiNALMo set new benchmarks across multiple RNA-related tasks. It achieved state-of-the-art results in secondary and tertiary structure prediction, surpassing previous models in accuracy and reliability. Additionally, RiNALMo demonstrated exceptional capabilities in predicting RNA-protein and RNA-RNA interactions, key for understanding cellular processes and developing RNA-based therapeutics.

RNA language modeling currently lacks a foundational model that can handle a wide range of tasks through a straightforward token-masking framework without depending on task-specific data or dedicated pre-processing. Such a model would excel at efficiently learning from the intrinsic structure of RNA sequences.

To move away from imposed biases, we propose a new tokenization strategy. Instead of relying on single nucleotides, codons, or static k-mers, which each bring arbitrary assumptions and fixed nucleotide groupings, we introduce a learnable tokenization scheme that adapts to capture sequence details at multiple levels of granularity. Paired with a BERT-like transformer optimized for contextual understanding, this approach achieves competitive or superior performance relative to larger, task-specific RNA models while substantially reducing parameter demands.
\section{Methods}

\subsection{ChaRNABERT architecture}

The ChaRNABERT (CRB) architecture is designed to be able to capture both fine-grained nucleotide details and broader contextual relationships efficiently, optimized for understanding the complex structures of RNA. At its core, CRB employs a modified Gradient-Based Subsequence Tokenization (GBST) \citep{tay2022charformer}, paired with a bidirectional BERT encoder \citep{devlin-etal-2019-bert}. This combination allows the model to dynamically identify and emphasize biologically relevant subsequences without the constraints of a predefined vocabulary. Simultaneously, it captures the long-range dependencies and bidirectional context crucial for accurately modeling RNA structures and functions.

\begin{figure*}
    \centering
    \includegraphics[width=\linewidth]{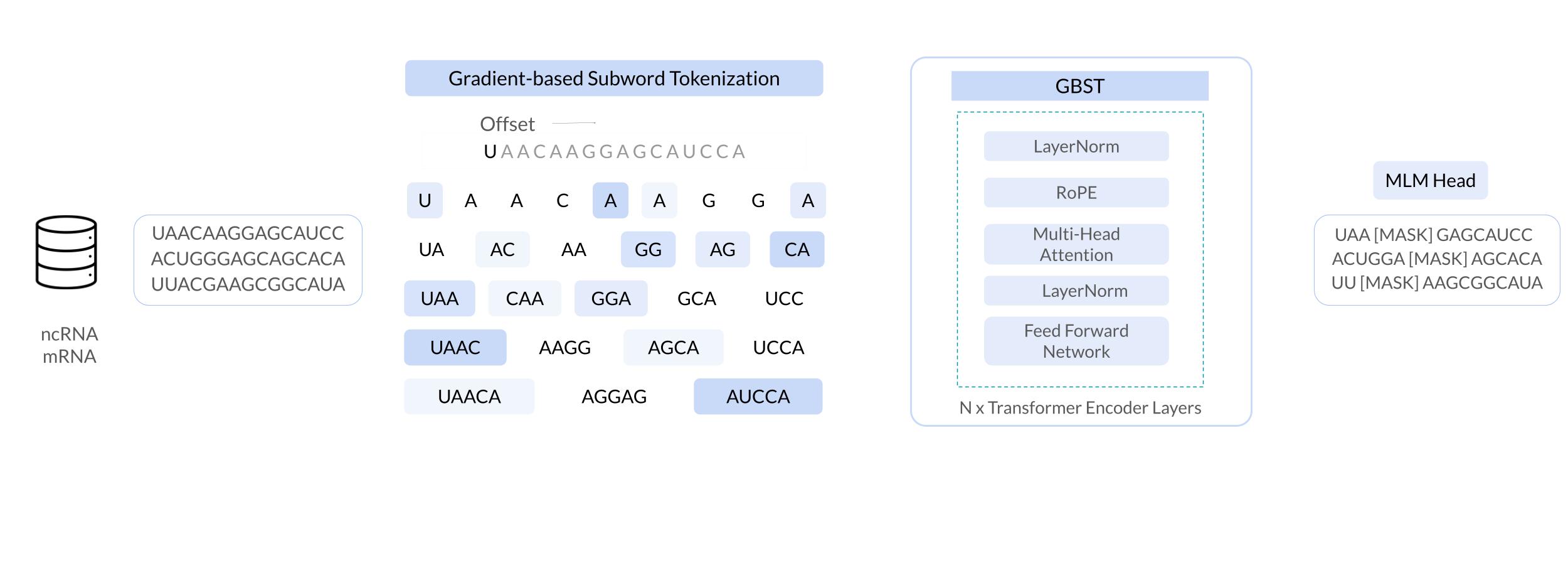}
    \caption{ChaRNABERT's architecture. We train our models utilizing two datasets: one with exclusively non-coding sequences and another that combines both coding and non-coding sequences. Through Gradient-Based Subsequence Tokenization (GBST), the model learns optimal tokenization patterns for RNA sequences. ChaRNABERT employs a standard BERT transformer encoder, accommodating an input context of up to 8,190 nucleotides, and is trained using a masked language modeling objective to capture sequence information.}
    \label{fig:enter-label}
\end{figure*}

\subsection{Character-level tokenization}

To effectively model subsequence information directly from nucleotide-level inputs, we employ a "soft" subword tokenization approach from character-level inputs. The original idea behind this approach is to allow the model to learn latent subsequence segmentations by dynamically selecting the most appropriate subsequence block at each character position during training procedure. 

This key idea is extended through the enumeration of offsets in the sequences in a sliding window manner, as to model the equivalent of open reading frames (ORFs). The learnable combination of both approaches allows us to dynamically select the best tokenization possible for each of the nucleotides and whether or not to take into account the local environment of the sequence. 

We also differ from the original implementation in the removal of the downsampling procedure, as single nucleotide resolution is highly desirable for several downstream applications.

\subsubsection{Constructing Candidate Subsequence Embeddings}
Given an input sequence of nucleotides, they are embedded as a tensor  $\mathbf{X} \in \mathbb{R}^{L \times d}$, where $L$ is the sequence length and $d$ is the nucleotide embedding dimension. To this individual representation GBST applies a one-dimensional depthwise convolution of kernel size equal to the maximum block size $M$, that acts as a smoothing operation, encouraging block level representations and allowing model to consider small shifts in the starting positions of blocks and the influence of non-adjacent nucleotides to some extent.
\begin{equation}
\tilde{\mathbf{X}} = \mathrm{1D\ DWConv}\left( \mathbf{X} \right),
\end{equation}

Afterwards, we generate candidate subsequence blocks by enumerating contiguous and overlapping spans of nucleotides up to a maximum block size $M$. For each block size $b$ (where $1 \leq b \leq M$), we construct subsequence blocks $\mathbf{X}_{b,i}$ starting at position $i$ by applying a pooling function over the embeddings of the nucleotides in the block:

\begin{equation}
    \tilde{\mathbf{X}}_{b,i} = F\left( \tilde{\mathbf{X}}_{i:i+b} \right),
\end{equation}

where $F: \mathbb{R}^{b \times d} \rightarrow \mathbb{R}^{d}$ is a non-parametric pooling function, in our case a sum pooling, that aggregates the embeddings within the selected subsequence into a single vector.

This procedure is also repeated for $b - 1$ offsets $o$ of $1$ for each of the block sizes in a sliding window manner. For example, for block size of one it obtains single nucleotides representations, whereas for block size two it is able to extract information from each pair of nucleotides starting from position 0 and position 1. This is continued up to the maximum block size $M$.

\begin{equation}
\tilde{\mathbf{X}}_{b,i,o} = F\left( \tilde{\mathbf{X}}_{i+o:i+b+o} \right),
\end{equation}

\subsubsection{Forming Latent Subsequence Representations}

To determine the most suitable subsequence block and offset at each nucleotide position, the approach introduces a scoring network $F_R: \mathbb{R}^{d} \rightarrow \mathbb{R}$. This network computes a scalar score $p_{b,i}$ for each candidate $\mathbf{X}_{b,i,o}$, reflecting the model's confidence in selecting that representation:

\begin{equation}
p_{b,i,o} = F_R\left( \tilde{\mathbf{X}}_{b,i,o} \right).
\end{equation}

We then compute a softmax over the scores for all representation sizes at position $i$, producing a probability distribution $P_i$ over the candidate blocks and offsets:

\begin{equation}
    P_i = \text{softmax}\left( [p_{1,i}, \, p_{2,i}, \, \dots, \, p_{M,i}] \right).
\end{equation}

This probabilistic weighting allows the model to softly select among the candidate representations based on their scores.

To enhance the model capabilities of capturing global context in the initial representation selection, we decided to incorporate the position-wise score calibration procedure from the original implementation. This layer computes a pseudo self-attention score between the different positions, encouraging the model to learn consensus among representation selection across the entire sequence. Specifically, updates are applied to the representation scores $P$ using a self-attention mechanism without additional projections:

\begin{equation}
    \tilde{P} = \text{softmax}\left( P P^\top \right) P,
\end{equation}

where $P \in \mathbb{R}^{L \times M}$ is the matrix of block probabilities, and $\tilde{P}$ is the consensus probability matrix.

The latent subsequence representation at position $i$ is obtained by computing a weighted sum of the candidate representation embeddings, using the probabilities from the modified probability matrix as weights:
\begin{equation}
    \hat{\mathbf{X}}_i = \sum_{b=1}^{M} \sum_{o=0}^{M-1} \hat{P}_{b,i,o} \, \tilde{\mathbf{X}}_{b,i,o}.
\end{equation}

This operation effectively allows the model to learn a soft subsequence segmentation, where each nucleotide position contributes to the final representation based on the likelihood of various block sizes and offsets. This soft selection mechanism ensures that the entire process is differentiable, enabling data-driven changes in the tokenization scheme during the training procedure and end-to-end training of the model.

Overall, by integrating GBST into our model, we leverage the strengths of subword representations while maintaining the flexibility, adaptability and resolution of nucleotide-level processing.

\subsection{Bidirectional BERT encoder}

As the main architecture we employ a BERT-based model with a few improvements. This model is a transformer encoder that enables bidirectional context learning through a self-attention mechanism and pre-training objectives. Each input token is mapped to an embedding and tokenized, combined with a positional encoding to maintain token order, and passed through multiple layers of the encoder. The core mechanism in each layer is \textit{multi-head self-attention} (multi-heads are omited from all equations for clarity), where for each token $i$, its attention with all tokens $j$ in the sequence is computed as:

\begin{equation}
\text{Attention}(Q, K, V) = \text{softmax}\left( \frac{QK^T}{\sqrt{d_k}} \right)V
\end{equation}

where $Q = XW_Q$, $K = XW_K$, and $V = XW_V$ are linear projections of the input $X$, i.e. token embeddings, and $d_k$ is the dimensionality of the keys/queries. 

We introduce a few other common architectural modifications, namely SwiGLU's non-linearities \citep{shazeer2020gluvariantsimprovetransformer}, Rotary Positional Encodings (ROPE) \citep{roformer}, Query Key Normalization (QKNorm) \citep{scalingvzit} and Flash Attention 2 \citep{dao2023flashattention2}.

\begin{equation}
    \text{SwiGLU}(x) = \sigma(xW_1) \odot \text{swish}(xW_2)
\end{equation}
where the Swish function is defined as:

\begin{equation}
    \text{swish}(x) = x \cdot \sigma(x)
\end{equation}

and $\sigma$ is the standard sigmoid non-linearity.
\begin{equation}
    \sigma(x) = \frac{1}{1 + e^{-x}}
\end{equation}

SwiGLU non-linearities are a combination of the Swish and Gated Linear Unit (GLU) non-linearities that have shown improved performance over its individual parts or classical functions like ReLU. 

We additionally move away from absolute positional encodings and introduce ROPE to our model, which has been shown to increase performance and length generalization capabilities in comparison with absolute and other relative positional encodings.
This approach is based upon a rotation mechanism, where positions in the sequence are represented as rotations in the embedding space.

\begin{equation*}\hspace*{-2.6cm}
f_{\{q,k\}}(x_m, m) = 
\begin{pmatrix}
\cos m\theta & -\sin m\theta \\
\sin m\theta & \cos m\theta
\end{pmatrix} \times
\end{equation*}

\begin{equation}\hspace*{+.5cm}
\times\begin{pmatrix}
W^{(11)}_{\{q,k\}} & W^{(12)}_{\{q,k\}} \\
W^{(21)}_{\{q,k\}} & W^{(22)}_{\{q,k\}}
\end{pmatrix}\times
\begin{pmatrix}
x_m^{(1)} \\
x_m^{(2)}
\end{pmatrix}
\end{equation}

with to the following values for $\theta$ to add the the long-term decay property between the relative positions to the positional encoding.

\begin{equation}
    \theta_i = 10000^{-2i/d_k}
\end{equation}

Moreover, to reduce the amount of training instabilities and loss spikes we decide to introduce QKNorm, effectively reducing the growth of the attention logits, which we found was a cause of instability in our training procedure.
This mechanism applies a LayerNorm (LN) to the output of the Query and Key linear transformations. 

\begin{equation}
   \text{softmax}\left[\frac{1}{\sqrt{d_k}}\text{LN}(XW^Q)(\text{LN}(XW^K))^T\right]
\end{equation}

Lastly, we include the hardware-aware Flash Attention 2 algorithm, allowing for an efficient increase of our context window and accelerated computation during training and inference.

\section{Pretraining}

\subsection{Masking strategies}

Typically, BERT’s pre-training uses a masked language model (MLM) objective, where a random subset of tokens is replaced with a special mask token. The model is trained to predict the original tokens based on both preceding and following context, forcing it to encode bidirectional information. The prediction of the masked token is computed as:

\begin{equation}\label{softmax_mask}
P(\text{token}_i | X_{\text{masked}}) = \text{softmax}(W_o h_i)
\end{equation}

where $h_i$ is the hidden state of token $i$ after passing through multiple self-attention layers, and $W_o$ is a learned output projection matrix. 

Learning from more complex corruption schemes can enhance a model's ability to capture complex patterns and dependencies, therefore we chose to incorporate the UL2 (Unifying Language Learning) \citep{tay2023ul} paradigm into our training regimen. UL2 is a pre-training framework that unifies various language modeling objectives to create a more versatile and robust language model.

It introduces a novel masking strategy that combines different types of denoising objectives. These include short-span masking (S-denoising), extreme-span masking (X-denoising), and retrieval-augmented masking (R-denoising). S-denoising is similar to BERT's MLM objective, where individual tokens or short spans are randomly masked within the input sequence, and the model learns to predict these masked tokens using bidirectional context. X-denoising involves masking longer contiguous spans of text, which forces the model to understand and reconstruct larger chunks of information, thereby enhancing its ability to handle longer dependencies. R-denoising trains the model in an autoregressive fashion, predicting future tokens based on past context, akin to models like GPT.

UL2 employs a shared g-masked token between the strategies to replace the masked spans, providing a unified way for the model to identify and reconstruct the missing information regardless of the span length. These strategies are selected based on a predefined sampling strategy, exposing the model to the different denoising strategies along the training process.

This process involves preparing the input by selecting a mode (S, X, or R) according to a series of specified probabilities and masking the input text accordingly. The model processes the masked input to generate hidden states for each token, and for the masked positions, it predicts the original tokens using the surrounding context. The training objective is to minimize the cross-entropy loss between the model's predictions and the actual masked tokens across all modes. Therefore the prediction for each masked token $i$ can be computed in the same way as the MLM objective (see Eq.\ref{softmax_mask}).

By leveraging UL2 masking, the model benefits from enhanced context understanding, versatility, and improved generalization. First, training on both short and long spans allows the model to comprehend and generate text over varying lengths, improving its understanding of context and long-range dependencies. Second, the combination of bidirectional and autoregressive objectives enables the model to perform well on a wide range of tasks.%, from understanding to generation.
Last, exposure to different types of denoising tasks helps the model generalize better to unseen data and tasks.

\subsection{RNA datasets}

For our study, we employed RNAcentral \citep{rnacentral} as the primary source for non-coding RNA sequences in our training dataset. RNAcentral is an extensive repository that consolidates non-coding RNA data from multiple expert databases, providing a unified and comprehensive resource. The dataset encompasses a diverse range of RNA families, including but not limited to: microRNAs (miRNAs), small nuclear RNAs (snRNAs), small nucleolar RNAs (snoRNAs), transfer RNAs (tRNAs), ribosomal RNAs (rRNAs), long non-coding RNAs (lncRNAs), Piwi-interacting RNAs (piRNAs), and small interfering RNAs (siRNAs). In total, RNAcentral contributed approximately 31 million non-coding RNA sequences to our dataset, offering a rich and varied collection for training our models.

To enable a comprehensive analysis that includes both non-coding and coding sequences, we expanded our dataset by incorporating coding sequences from RefSeq (Reference Sequence) database at the National Center for Biotechnology Information (NCBI) \citep{refseq-rn}. Specifically, we added 31 million coding sequences to our dataset. This augmentation resulted in a balanced and extensive dataset comprising both non-coding and coding sequences, which is crucial for training robust models capable of distinguishing between the two types.

\subsection{Model sizes}

To assess the impact of model size on performance and explore scalability in RNA sequence analysis, we trained models with parameter counts mostly aligned with the main ESM models, specifically developing models with approximately 8M, 33M, 50M, 100M, 150M, and 650M parameters. 

Our exploration of scaling effects, both with and without GBST, involved training these models to examine the combined impact of character-level tokenization with the BERT encoder. This investigation not only focuses on MLM/UL2 loss performance but also evaluates the models’ effectiveness on downstream tasks and generalization capabilities.

Despite we investigate scaling parametrically (see Section \ref{sec:scaling_parametric}), we chose to train models at these specific sizes as a baseline, even though it may not represent the optimal approach for all scenarios. This decision allows us to better analyze the interaction between model size and GBST. Our objective is to understand how these factors influence not only loss metrics but also broader performance across downstream applications. 

For a thorough analysis, all model sizes were trained on two datasets: 31 million non-coding RNA sequences from RNAcentral and the combination of this dataset and the 31 million coding sequences from RefSeq.

All the models are trained in BF16 precision using Distributed Data Parallel with the DeepSpeed \citep{deepspeed} and ZeRO \citep{zero} frameworks for maximum optimization of computational resources.
\begin{table}[h!]
    \centering
    
    \begin{tabular}{rccccc}
        \toprule
        \textbf{Parameters} & \textbf{num\_layers} & \textbf{d\_model}  & \textbf{num\_heads}   \\
        \midrule
        8M   &  6 & 320   & 20 \\
        33M  & 12 & 480   & 20 \\
        50M  & 15 & 500   & 20 \\
        100M & 23 & 600   & 20 \\
        150M & 30 & 640   & 20 \\
        650M & 33 & 1280  & 20 \\
        \bottomrule
    \end{tabular}
    \caption{Dimensions by model parameters of ChaRNABERT models.}
    \label{model_params}
    \vskip 0.1in
\end{table}

\section{Experiments}
In this section, we show several analyses on the factors that influence the performance of ChaRNABERT in order to identify the settings that provide the best performance of the model as well as to identify the optimal trade-off between performance and computational cost. 

Advances in large language models (LLMs) have been driven by scaling up parameters, enhancing their applications in natural language processing (NLP), as detailed by the studies of \citep{kaplan2020scaling} and \citep{hoffmann2022training}. While scaling studies have also progressed in fields like pLMs \citep{pLMsscaling}, with research investigating model size effects, a comprehensive analysis of LLM scaling applied to RNA remains underexplored. Works such as RiNALMo \citep{rinalmo} have examined performance differences under variations of the parameter count, yet a detailed investigation of RNA-specific scaling laws is still absent. This analysis aims to bridge that gap, using ChaRNABERT as an initial model to guide future large-scale RNA LLM research.

First, we explore how various learning rates and context window sizes affect ChaRNABERT's performance across model sizes, aiming to understand the impact of key hyperparameters. Next, we assess the model's efficiency with datasets of different sizes to evaluate the effects of data scaling. Finally, we analyze computational efficiency by measuring the impact of increased floating-point operations (FLOPs) on model improvements. Throughout, we follow \citep{hoffmann2022training}'s scaling principles and compare tokenization strategies, highlighting the performance gains of using GBST over embeddings alone.

\subsection{Impact of learning rate and context window}

We analyzed the impact of using different learning rates and context window sizes on three models of different sizes: 5 million, 50 million, and 100 million parameters, using a dataset composed of 31 million non-coding sequences extracted from RNACentral \citep{rnacentral}. 

For each model size, three learning rates were tested, selected on the basis of the number of parameters in the model. Specifically, for the 5M and 50M parameter models, the learning rates tested were 5e-4, 1e-4, and 5e-5. For the 100M parameter model, the learning rates tested were 1e-4, 5e-5, and 1e-5. We choose different learning rates per model size in order to avoid training instabilities. The aforementioned tests permitted an evaluation of the influence of the learning rate on convergence.

In general, it was observed that in smaller models (Table \ref{tab:5Mlr}), such as the 5 million parameter model, higher learning rates, 5e-4 or 1e-4, achieved a slightly lower loss compared to lower rates like 5e-5. Nevertheless, these gains, while present, were not substantial. In the middle models, such as the 50M parameter model (Table \ref{tab:50Mlr}), learning rates such as 1e-4 or 5e-5 achieved lower losses compared to higher learning rates like 5e-4 which resulted in high instability. For larger models, such as 100M model (Table \ref{tab:100Mlr}), higher learning rates, 1e-4 and 5e-5, similarly resulted in a lower loss compared to 1e-5. However, upon examining the convergence curves, lower learning rates helped to avoid instability in mid and large-sized models.

\begin{table}[h!]
\centering
\renewcommand{\arraystretch}{1.2}
\begin{adjustbox}{width=\columnwidth}
\begin{tabularx}{\columnwidth}{l >{\raggedleft\arraybackslash}X >{\raggedleft\arraybackslash}X >{\raggedleft\arraybackslash}X}
\toprule
\textbf{Model | Learning Rate} & \textbf{5e-4} & \textbf{1e-4} & \textbf{5e-5} \\ 
\midrule
EM         & 0.508  & 0.580  & 0.639        \\ 
GBST       & 0.498  & 0.537  & 0.581        \\ 
\bottomrule
\end{tabularx}
\end{adjustbox}
\caption{Comparison of Final Loss by Learning Rate - Model Size 5M}
\label{tab:5Mlr}
\end{table}

\begin{table}[h!]
\centering
\renewcommand{\arraystretch}{1.2}
\begin{adjustbox}{width=\columnwidth}
\begin{tabularx}{\columnwidth}{l >{\raggedleft\arraybackslash}X >{\raggedleft\arraybackslash}X >{\raggedleft\arraybackslash}X}
\toprule
\textbf{Model | Learning Rate} & \textbf{5e-4} & \textbf{1e-4} & \textbf{5e-5} \\ 
\midrule
EM         & 3.516 & 0.437   & 0.469        \\ 
GBST       & 19.52  & 0.434  & 0.450        \\ 
\bottomrule
\end{tabularx}
\end{adjustbox}
\caption{Comparison of Final Loss by Learning Rate - Model Size 50M}
\label{tab:50Mlr}
\end{table}

\begin{table}[h!]
\centering
\renewcommand{\arraystretch}{1.2}
\begin{adjustbox}{width=\columnwidth}
\begin{tabularx}{\columnwidth}{l >{\raggedleft\arraybackslash}X >{\raggedleft\arraybackslash}X >{\raggedleft\arraybackslash}X}
\toprule
\textbf{Model | Learning Rate} & 
\textbf{1e-4} &  \textbf{5e-5} & \textbf{1e-5} \\ \midrule
EM         & 0.446  & 0.465  & 0.640        \\ 
GBST       & 0.454  & 0.460  & 0.577        \\
\bottomrule
\end{tabularx}
\end{adjustbox}
\caption{Comparison of Final Loss by Learning Rate - Model Size 100M}
\label{tab:100Mlr}
\end{table}

\vspace{0.3cm}

\begin{figure}[h!]
    \centering
    \includegraphics[width=\linewidth]{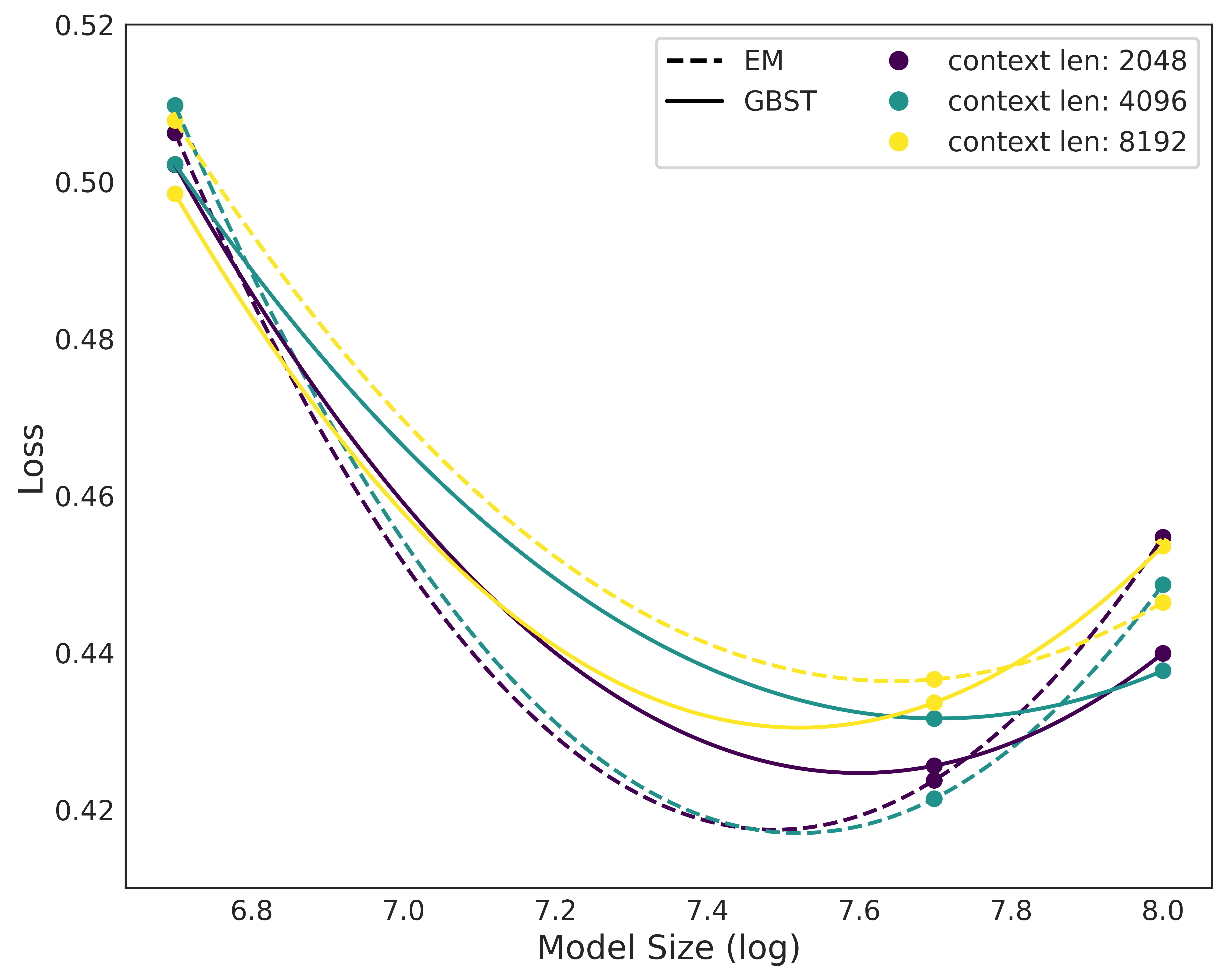}
    \vskip 0.1in
    \caption{Performance comparison with different context windows. We present the final exponential moving average (EMA) loss results for two tokenisation methods, EM (dashed lines) and GBST (solid lines), evaluated in different model size configurations, which are plotted in logarithmic scale on the X-axis. Each colour indicates a different context window length (2048, 4096 and 8192).}
    \label{fig:seq_len}
\end{figure}

To explore how context window size affects performance, tests were conducted with window sizes of 8192, 4096, 2048 (Figure \ref{fig:seq_len}). For RNA language models, the context window is especially relevant due to the biological nature of the sequences processed by these models. Interactions, such as base pairing and secondary structures, can be scattered throughout the sequence, making it essential that the model is able to capture a context wide enough to identify relevant patterns.

The results in Figure \ref{fig:seq_len}, reveal a consistent trend for both tokenization strategies: single nucleotide embedding (EM) and GBST. The curves generally follow a U-shaped pattern, indicating that for each sequence size there is an optimal point at which the model reaches its minimum loss. After this point, further increase in model size result in diminishing returns. This saturation point appears to be around 30 to 50 million parameters, beyond which there is no significant improvement in performance, and the loss tends to increase slightly.

One notable aspect of these results is that, despite the existence of some differences between the context window sizes, these are not pronounced enough to conclude that one size is clearly superior to another in terms of performance. Nevertheless, is noteworthy that, in models with EM tokenization, smaller window sizes (2048 and 4096) tend to achieve lower final losses compared to larger windows such as 8192, suggesting that a smaller window might be more effective in certain cases. This indicates that, although the size of the context window has some impact, this seems to be more limited, especially when GBST tokenization is used.

\subsection{Varying token counts}\label{section_token_counts}

We also aimed to identify the impact of token count in model performance. We retained the three model sizes, 5M, 50M and 100M parameters, and generated datasets of varying sizes: 15M, 66M, 100M and 150M sequences which by the average number of tokens per sequence correspond to 2.4B, 10.9B, 16.5B, and 24.8B tokens respectively.

\begin{figure}[h!]
    \centering
    \includegraphics[width=\linewidth]{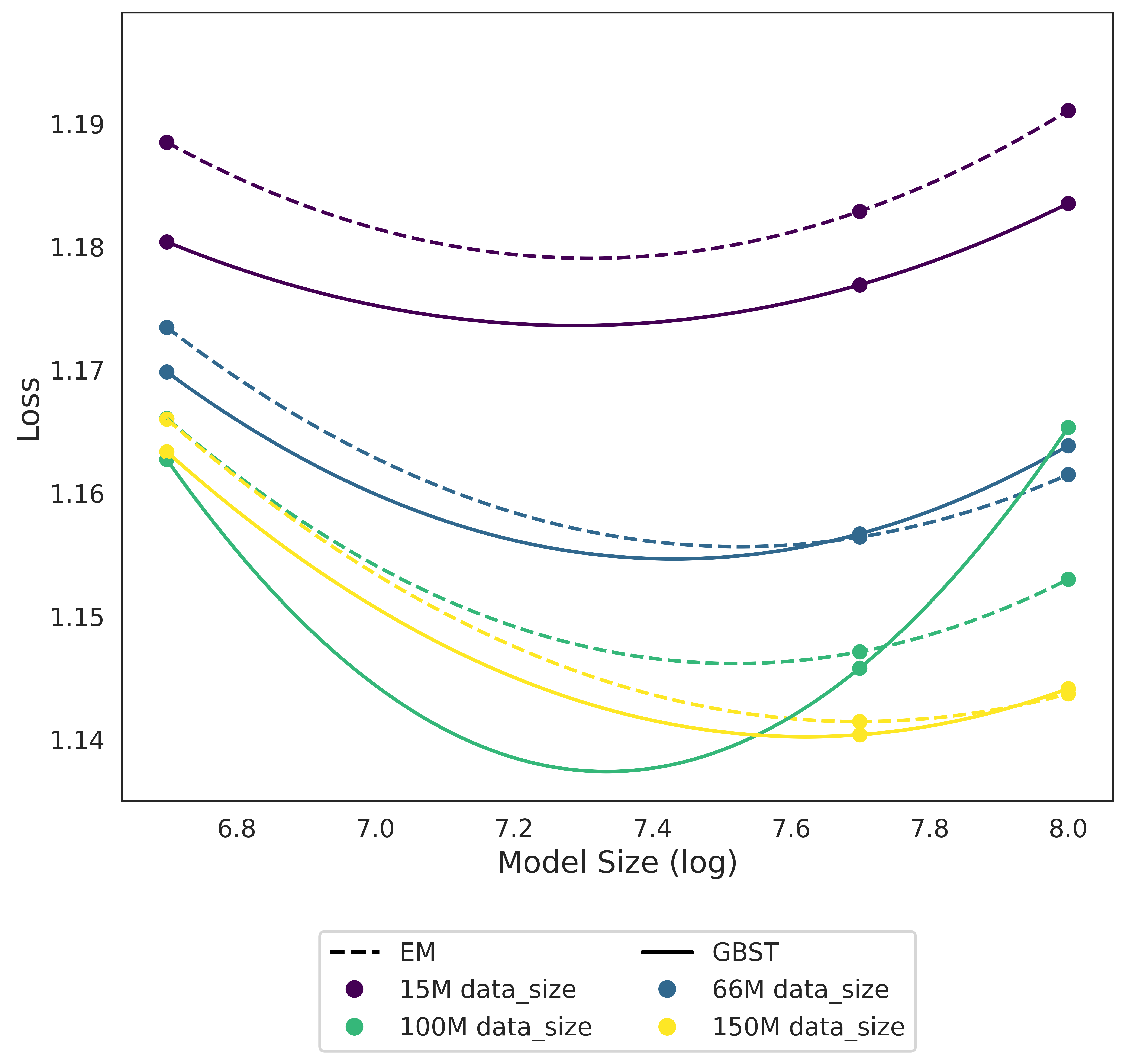}
    \vskip 0.1in
    \caption{Performance comparison across data sizes. The figure displays the final EMA loss for two tokenization methods, EM (dashed lines) and GBST (solid lines), evaluated across various model sizes (plotted in logarithmic scale on the X-axis). Different colors represent distinct data sizes (15M, 66M, 100M, and 150M of sequences). }
    \label{fig:mars}
\end{figure}

Since the amount of non-coding sequences is limited, we included both coding and non-coding, extracted from the MARS \citep{mars} sequence database. The sequences were randomly selected. In Figure \ref{fig:mars}, the loss obtained per model size and per dataset for both tokenization methods, GBST and EM, is shown.

From the results displayed in Figure \ref{fig:mars}, it can be observed that the size of the dataset has a visible but not substantial impact on the overall performance of the model. In both GBST and EM tokenization experiments,  the losses decreased slightly as the dataset size increased, but these improvements were not substantial enough to drastically improve the performance of the models. 
Furthermore, for both tokenization techniques, a clear U-shaped trend is still observed in the plots, suggesting that model performance can saturate at around 30M to 50M parameters, regardless of the size of the dataset used. 

Based on the analysis conducted, it is clear that while learning rate and context window size have some impact on performance, the factor that most significantly affects ChaRNABERT's performance is the size of the model. Larger models tend to perform better up to a certain point, after which performance gains become marginal, and in some cases, loss even increases slightly. In the following section, we will further explore this by conducting a detailed study on model scaling, with the aim of understanding how the number of parameters influences performance and identifying the optimal scaling strategies for ChaRNABERT.

\begin{figure}[h!]
    \centering
    \includegraphics[width=\linewidth]{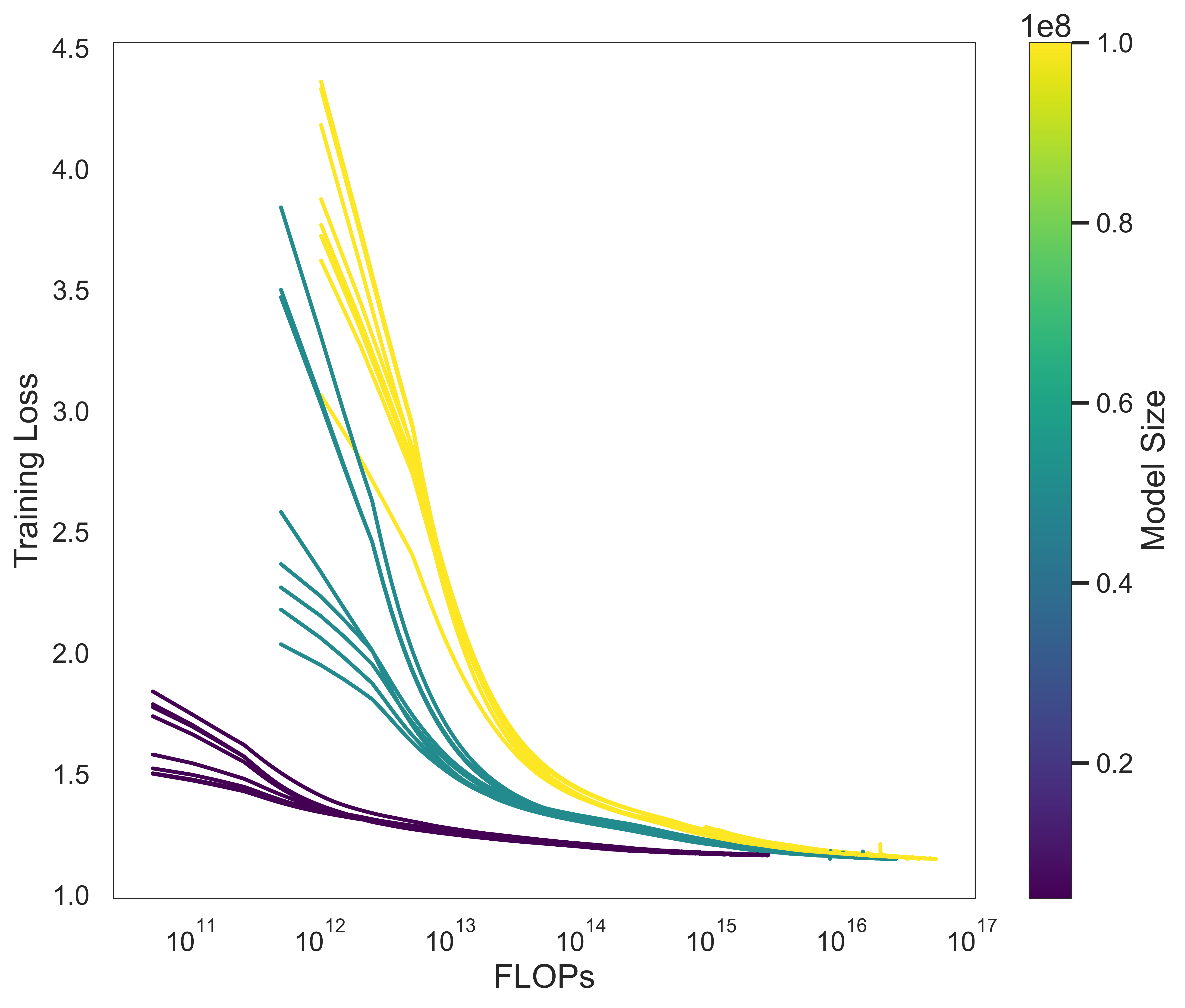}
    \vskip 0.1in
    \caption{FLOPs vs. training loss across model sizes. We show the relationship between computational cost (FLOPs, on a logarithmic scale, X-axis) and training loss (Y-axis) for models of varying sizes.}
    \label{fig:flopsvsloss}
\end{figure}

\subsection{Increasing model size} \label{sec:scaling_parametric}

We follow \citet{hoffmann2022training} to fit a parametric loss function. Using the data collected during the experimental phases, we fit the power laws to establish the relationships \( N_{opt} \propto C^a \) and \( D_{opt} \propto C^b \), where \( N \) is the model size, \( D \) is the number of tokens, and \( C \) represents the computational budget in FLOPs\footnote[1]{FLOPs computation is equivalent to the one defined in \citet{hoffmann2022training} and can be found at Appendix \ref{flops}.}. The exponents \( a \) and \( b \) were determined based on the fitted parameters, of the scaling law:

\begin{equation}\label{hoff}
   \hat{L}(N, D) \triangleq E + \frac{A}{N^\alpha} + \frac{B}{D^\beta}
\end{equation}

Here, \( N \) is the number of model parameters, \( D \) is the size of the data set (in tokens), and \( E \) captures the natural entropy of the text (ideal loss). The terms with \( A \) and \( B \) reflect the deviation of the model from the ideal loss, due to the limited size of the model and data. The exponents \( \alpha \) and \( \beta \) determine the impact of model and dataset size on the loss\footnote[2]{Parameter counts for our models are defined at Appendix \ref{param:comp}.}.

We fit the expression in Eq. \ref{hoff} following \citep{hoffmann2022training}\footnote[3]{Detailed procedure can be found at Appendix \ref{scaling}.}. In particular, we utilized the most optimal results obtained for each model size from the experiment described in Section \ref{section_token_counts}. This analysis was intended to capture the effects of both model size and dataset size on the model performance.

We found that for both GBST and EM, the final loss decreases predictably as the model parameters increase (Figure \ref{fig:scaling_law}), following a general trend of improvement with larger models. However, in both cases, improvements plateau or even cease to be significant once a certain parameter threshold is exceeded, particularly around 30 to 50 million parameters.

\begin{table*}[h!] 
\centering

\begin{subtable}[h!]{\textwidth}
\centering
\begin{tabular}{@{} l *{7}{>{\centering\arraybackslash}p{1.8cm}} @{}}
\toprule
\textbf{Task}   & \textbf{SSP} & \textbf{CMP} & \textbf{DMP} & \textbf{SSI} & \textbf{SPL} & \textbf{APA} & \textbf{NcRNA} \\ \midrule
\textbf{Metric} & F1 & P@L$R^{}$ & $R^{2}$ & F1 & $R^{2}$ & ACC \\ \midrule
Baseline & 0.59 $\pm$ 0.00 & \textbf{0.60} $\pm$ \textbf{0.01} & 0.45 $\pm$ 0.00 & 0.38 $\pm$ 0.00 & -- & 0.67 $\pm$ 0.01 & 0.89 $\pm$ 0.00 \\ 
RNA LM & 0.69 $\pm$ 0.01 & \textbf{0.60} $\pm$ \textbf{0.07} & 0.56 $\pm$ 0.00 & 0.42 $\pm$ 0.00 & -- & 0.73 $\pm$ 0.85 & 0.97 $\pm$ 0.00\\ 
RiNALMo & \textbf{0.72} $\pm$ \textbf{0.01} & 0.49 $\pm$ 0.06 & 0.59 $\pm$ 0.04 & 0.39 $\pm$ 0.01 & \textbf{0.96} $\pm$ 0.01 & \textbf{0.82} $\pm$ \textbf{0.01} & \textbf{0.98} $\pm$ \textbf{0.01} \\ 
CRB-8M & 0.61 $\pm$ 0.06 & 0.54 $\pm$ 0.03 & 0.59 $\pm$ 0.03 & \textbf{0.43} $\pm$ \textbf{0.00} & 0.93 $\pm$ 0.00 & \textbf{0.83} $\pm$ \textbf{0.01} & 0.95 $\pm$ 0.00 \\ 
CRB-33M & 0.52 $\pm$ 0.15 & 0.56 $\pm$ 0.02 & 0.60 $\pm$ 0.04 & \textbf{0.43} $\pm$ \textbf{0.00} & 0.93 $\pm$ 0.00 & \textbf{0.82} $\pm$ \textbf{0.00} & 0.95 $\pm$ 0.00\\ 
CRB-50M & 0.66 $\pm$ 0.06 & \textbf{0.59} $\pm$ \textbf{0.01} & \textbf{0.61} $\pm$ \textbf{0.02} & \textbf{0.43} $\pm$ \textbf{0.00} & 0.94 $\pm$ 0.00 & \textbf{0.83} $\pm$ \textbf{0.00} & 0.96 $\pm$ 0.01 \\ \bottomrule
\end{tabular}

\end{subtable}

\addtocounter{table}{-1} % Resets numbering for subtables

\vspace{0.5cm}

\begin{subtable}[h!]{\textwidth}
\centering
\begin{tabular}{@{} l *{6}{>{\centering\arraybackslash}p{2.167321cm}} @{}}
\toprule
\textbf{Task}   & \textbf{Modif} & \textbf{MRL} & \textbf{VDP} & \textbf{PRS} & \textbf{CRI-On} & \textbf{CRI-Off} \\ \midrule
\textbf{Metric} & AUC & $R^{2}$ & MCRMSE & $R^{2}$ & SC \\ \midrule
Baseline & \textbf{0.95} $\pm$ \textbf{0.01} & 0.84 $\pm$ 0.00 & 0.33 $\pm$ 0.00 & 0.55 $\pm$ 0.01 & 0.27 $\pm$ 0.01 & \textbf{0.12} $\pm$ \textbf{0.00} \\ 
RNA LM & \textbf{0.95} $\pm$ \textbf{0.00} & 0.85 $\pm$ 0.00 & 0.31 $\pm$ 0.00 & \textbf{0.58} $\pm$ \textbf{0.01} & 0.35 $\pm$ 0.00 & 0.05 $\pm$ 0.01 \\ 
RiNALMo & 0.76 $\pm$ 0.09 & 0.86 $\pm$ 0.01 & \textbf{0.23} $\pm$ \textbf{0.01} & 0.47 $\pm$ 0.02 & 0.39 $\pm$ 0.07 & 0.01 $\pm$ 0.04 \\ 
CRB-8M & 0.94 $\pm$ 0.00 & 0.86 $\pm$ 0.02 & 0.25 $\pm$ 0.00 & 0.52 $\pm$ 0.01 & 0.37 $\pm$ 0.01 & 0.11 $\pm$ 0.00\\ 
CRB-33M & 0.94 $\pm$ 0.00 & \textbf{0.90} $\pm$ \textbf{0.00} & 0.25 $\pm$ 0.01 & 0.52 $\pm$ 0.01 & \textbf{0.39} $\pm$ \textbf{0.01} & 0.11 $\pm$ 0.00 \\ 
CRB-50M & \textbf{0.95} $\pm$ \textbf{0.00} & \textbf{0.90} $\pm$ \textbf{0.00} & 0.25 $\pm$ 0.00 & 0.45 $\pm$ 0.01 & 0.35 $\pm$ 0.00 & 0.11 $\pm$ 0.01 \\ \bottomrule
\end{tabular}
\end{subtable}

\caption{Tasks and their associated metrics, with model performance values across six different models. Baseline corresponds to the best performing LSTM, CNN, or ResNet as stated in BEACON. RNA LM corresponds to the best performing RNA language model in BEACON. We report performance of ChaRNABERT model of sizes 8M, 33M, and 50M parameters. Average performances and standard deviations were computed over five independent runs. \textbf{Bold} measures correspond to the best performing model(s) for the given tasks under a Bonferroni-corrected t-test (p-value $<$ 0.05).}
\label{tab:tasks_metrics}
\end{table*}

From the obtained parameters we derived the power-law exponents for model size and token count as a function of compute \( N_{opt} \propto C^{0.2279} \) and \( D_{opt} \propto C^{0.7720} \). A key finding is that the optimal model size scales sublinearly with the compute budget. Specifically, the relationship \(N_{opt} \propto C^{0.2279}\) indicates that the model size grows at a slower rate than the compute budget. In contrast, the optimal number of training tokens scales superlinearly with compute. The relationship \(D_{opt} \propto C^{0.7720}\) shows that, as compute grows, the number of training tokens increases more rapidly than the model size.

We observe that model performance improves rapidly with increased compute at first, but after a certain threshold, approximately \(10^{16}\) FLOPs, the improvements in training loss begin to plateau (Figure \ref{fig:flopsvsloss}). This trend suggests that, while scaling up model size and compute provides significant early gains, we rapidly find a model size as a point where increasing parameter count yields diminishing returns.

\section{Assessing CRB's performance in downstream applications}

\subsection{BEACON benchmark}

\textbf{Implementing BEACON benchmark.} We leverage BEACON (BEnchmArk for COmprehensive RNA tasks and language models) to assess the performance of the CRB models in downstream tasks \citep{ren2024beacon}. BEACON is the first comprehensive benchmark designed to evaluate deep learning methods for RNA analysis, encompassing 13 tasks across structural analysis, functional studies, and engineering applications. BEACON evaluates both traditional models such as CNNs, ResNets, and LSTMs, as well as advanced RNA foundation models like RNA-FM and RNA-BERT.

\begin{figure}[htbp]
    \centering
    \includegraphics[width=\linewidth]{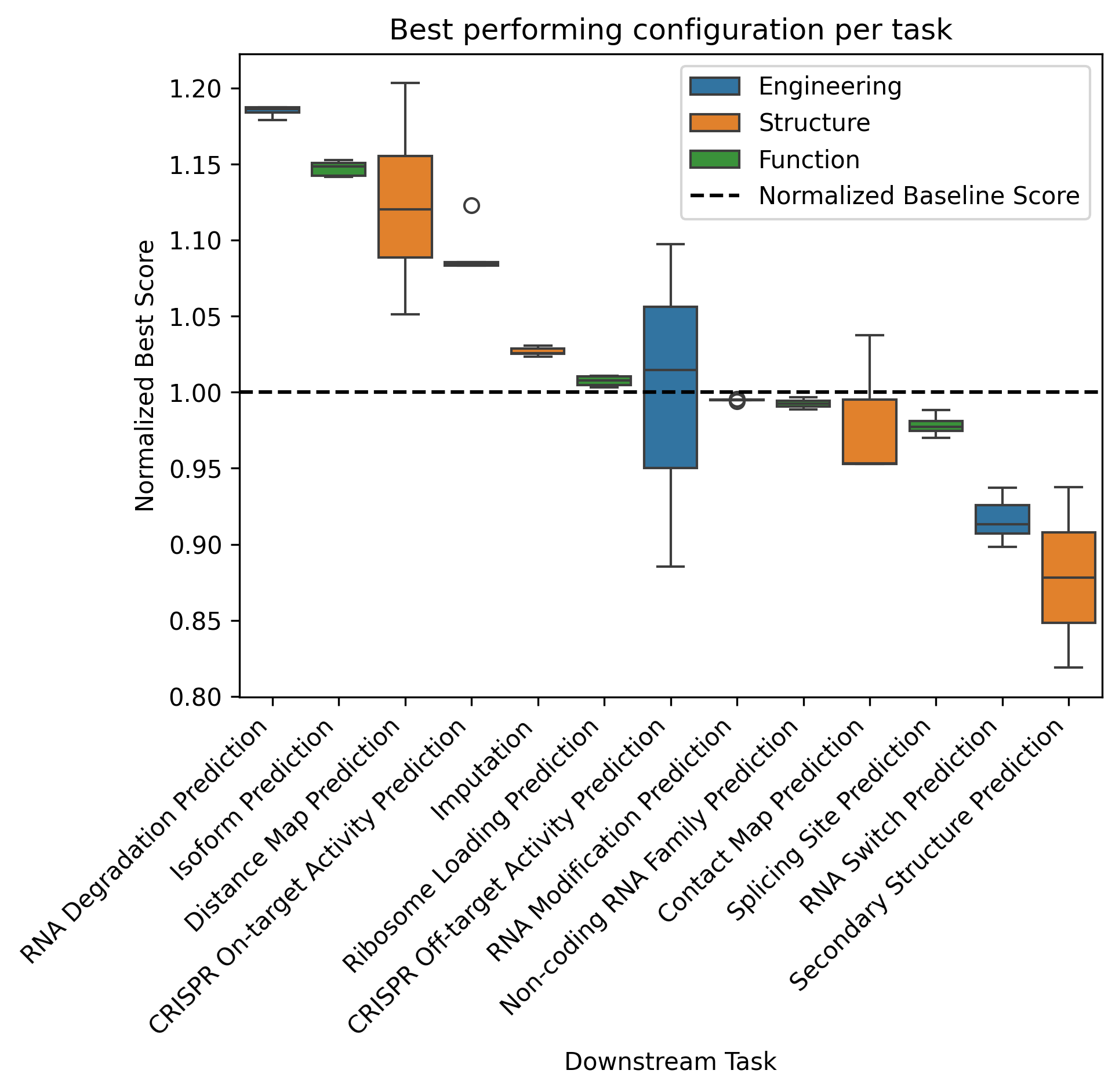}
    \vskip 0.1in
    \caption{Best performance achieved by CRB relative to the corresponding baseline for each downstream task.}
    \label{fig:best_scores}
\end{figure}

For each downstream task in the BEACON benchmark, we integrated a head module and trained it alongside the CRB architecture. The choice of the head module was based on the best benchmark available, either from BEACON or RiNALMo, depending on the type of task. ChaRNABERT successfully matched or exceeded the reference baselines for most downstream tasks, as shown in Figure \ref{fig:best_scores} and Table \ref{tab:tasks_metrics}. Multiple training runs were conducted for each task. 
%We fine-tuned models of all sizes, starting from the checkpoints generated during pre-training, including the best-performing checkpoint and individual epoch checkpoints.
We fine-tuned models of different sizes, dataset compositions and masking strategies, starting from the checkpoint corresponding to the first epoch of pre-training. A more detailed explanation of the implementation of BEACON's downstream tasks can be found at Appendix \ref{beacon_appendix}. For a more fair comparison, we add RiNALMo’s 650M model to the BEACON benchmark.

\textbf{CRB performs competitively on all tasks compared with the best RNA LM.} In Figure \ref{fig:best_scores} we show the performance of the best model configuration for each downstream task relative to the goal metric, obtained from BEACON benchmark or RiNALMo. For downstream tasks where the performance metric had to be maximized, the relative score was computed as the ratio between the best score obtained by CRB and the baseline score. For tasks where the metric aligns with the optimization objective (i.e., ranging from 0 to $+\infty$, where 0 is the optimal value), the relative score improvement was computed as the difference in scores relative to the baseline.

In structural prediction tasks, CRB models demonstrate competitive performance, particularly in mapping spatial relationships essential related to RNA’s tertiary structure. In Distance Map Prediction, CRB-50M surpasses all models, effectively capturing RNA spatial patterns. Similarly, in Structural Score Imputation, CRB-33M and CRB-50M outperform other models, showcasing their precision in reconstructing complex structural information despite missing data points. However, in Secondary Structure Prediction, RiNALMo, with its larger parameter count and capitalizing over a structure focused training, slightly outperforms CRB, indicating that while CRB captures structural features effectively, further optimization may enhance its capabilities in this area.

We observe that, for functional prediction tasks, CRB models excel in discerning biologically significant RNA sequence patterns. An example, APA Isoform Prediction, CRB-8M significantly outperforms other models, precisely predicts the usage ratio of the proximal polyadenylation site (PAS). Also, the CRB-50M model performs on par with RiNALMo, effectively predicting translation efficiency across different mRNA sequences. Nonetheless, in Non-coding RNA Function Classification and Splice Site Prediction, RiNALMo slightly edges out CRB, suggesting that for tasks involving nuanced ncRNA classifications, RiNALMo’s architecture and training may better accommodate diverse functional features. 

\begin{figure}[h!]
    \centering
    \includegraphics[width=\linewidth]{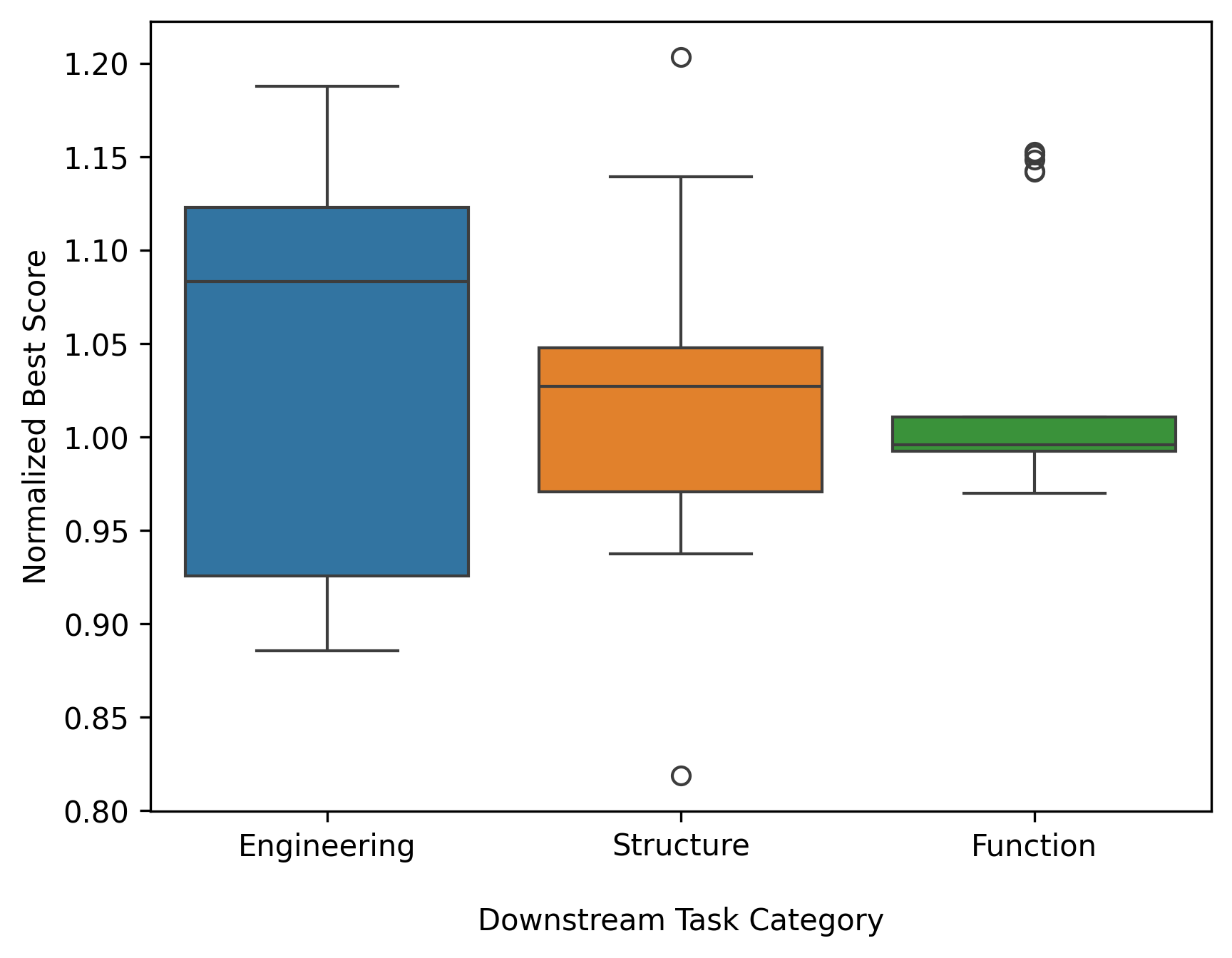}
    \vskip 0.1in
    \caption{Best performance achieved by CRB relative to the corresponding baseline for each downstream task category.}
    \label{fig:best_scores_by_category}
\end{figure}

In the engineering category, CRB models prove competitive, particularly in RNA vaccine degradation prediction, where the CRB-8M greatly outperforms reference models from the BEACON benchmark and achieves similar performance to RiNALMo. In CRISPR On-Target Prediction, CRB-33M matches top performance, revealing the model’s promise in guiding genome editing applications. However, for CRISPR Off-Target Prediction, a baseline model achieves higher accuracy than both CRB and RiNALMo, suggesting that simpler models may more effectively avoid overfitting, particularly in capturing off-target effects that may benefit from targeted training approaches. However, it is worth noticing that the performance of CRB outperforms any other RNA LM included in this study by a large margin on this particular task. 

Overall, we emphasize CRB's capacity to achieve competitive or state-of-the-art results with significantly fewer parameters than other models, all without requiring specialized training. This efficiency highlights the power of character-level tokenization in delivering high performance with reduced computational demands. A complete compilation of results over BEACON can be found in Table \ref{tab:tasks_metrics}.

\textbf{Certain tasks benefit from scaling but overall impact is small.} Figure \ref{fig:best_scores_by_category} presents boxplots illustrating the distribution of best scores across task categories. Notably, these distributions do not exhibit means significantly different from one, suggesting substantial variability in performance depending on the downstream task. This variability appears to lack a consistent pattern linked to task category, indicating that performance may be largely task-specific.

\begin{figure}[h!]
    \centering
    \includegraphics[width=\linewidth]{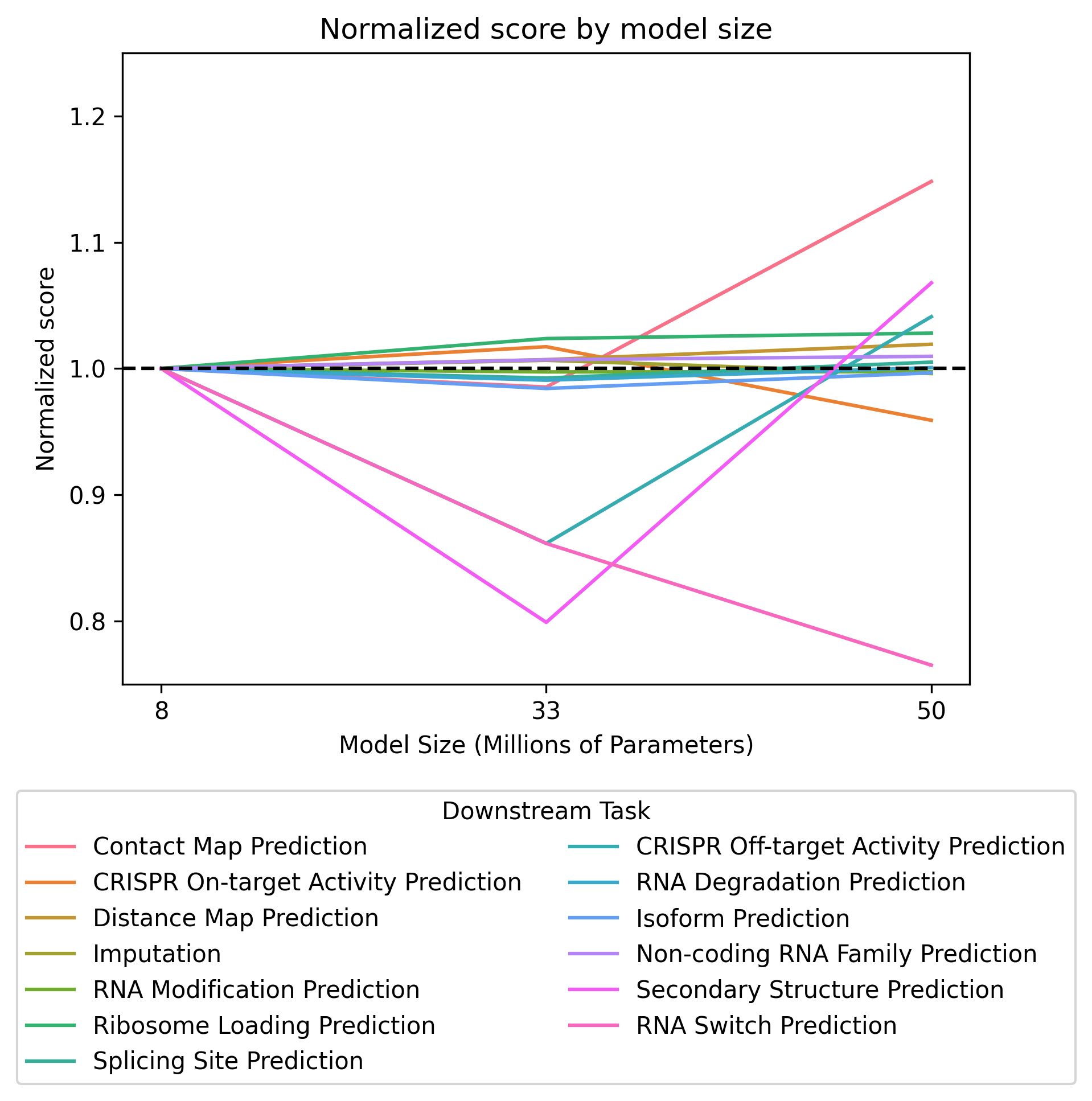}
    \vskip 0.1in
    \caption{Best performance achieved by CRB relative to the smallest model size for each downstream task.}
    \label{fig:scores_by_size}
\end{figure}

We also analyzed how model performance evolved across different downstream tasks and categories in relation to model size (Figure \ref{fig:scores_by_size}). With this purpose, we normalized the performance of different model configurations with respect to the mean performance of the smallest model. For most tasks the impact of increasing the model size from 8 to 50 million is relatively small. However, there are certain instances where the scale of the impact is large enough to be taken into account. This is the case for contact map prediction, where increasing model size from 33 to 50 million parameters causes an increase in performance of $\sim$15\%. An interesting case is the CRISPR off-target prediction task, where the model with 33 million parameters performs the worst, with significantly lower accuracy compared to models with 8 million and 50 million parameters.

\begin{figure}[h!]
    \centering
    \includegraphics[width=\linewidth]{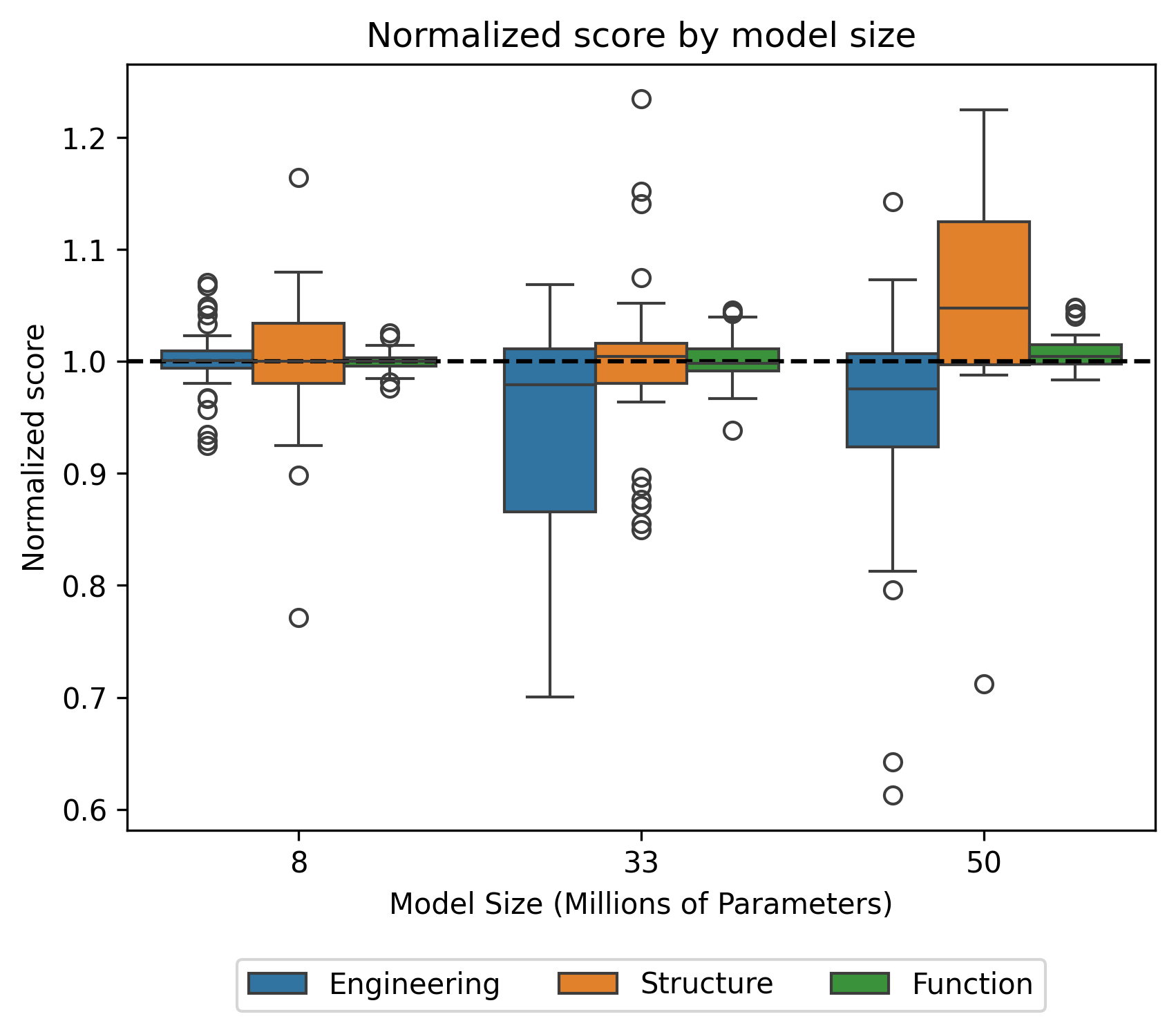}
    \vskip 0.1in
    \caption{Best performance achieved by CRB relative to the smallest model size for each downstream task grouped by task category.}
    \label{fig:scores_by_size_by_category}
\end{figure}

We further examined the impact of model size by grouping tasks into categories (Figure \ref{fig:scores_by_size_by_category}). For models between 8 and 50 million parameters, there were no statistically significant changes in mean performance across categories. This could reflect the varied influence of model size on different tasks within each group. However, we observed greater performance variability, particularly in engineering and structural prediction tasks, suggesting that model size may have a more nuanced effect in these areas.

\subsection{Extending the benchmark tasks for RNA}\label{extending-bechmark}

Given the numerous possible downstream applications of foundational language models, we wanted to benchmark the performance of our model across a broader range of tasks to explicitly identify their strengths and weaknesses. To this end, we leveraged existing databases to construct four new datasets that probe the limits of the models' capacity to generalize from sequence data to biologically relevant applications.

% \subsubsection{RNA-RNA interaction prediction}

% \textbf{ENCORI database}. We used the ENCORI database \citep{li2014starbase} to prepare a dataset consisting of experimentally verified intermolecular RNA-RNA contacts. First, we filtered out the lowest-confidence entries corresponding to a single read. From the remaining data, we sampled genomic sequences containing the entire identified interacting region together with genomic context of length 175-225 (drawn for each sequence from the uniform distribution). The task then is to predict the base-pairing (interaction) matrix between the two sequences, including non-continuous pairing and non-canonical base pairs, with a minimum of 9 base pairs forming the complementary region.

% \begin{table}[h!]
% \centering
% \begin{tabular}{lcc}
% \textbf{Model} & \textbf{Performance} \\ \hline
% LSTM        & 0.406 $\pm$  0.003 \\ 
% CNN        &  0.458 $\pm$  0.008  \\ 
% RiNALMo        & 0.532 $\pm$ 0.015  \\ 
% ChaRNABERT     & \textbf{0.542} $\pm$ \textbf{0.016} \\ \hline
% \end{tabular}
% \caption{F1 performance in the ENCORI derived dataset. \textbf{Bold} measures indicate the best model for a dataset under a t-test with p-value $<$ 0.05.}
% \label{table:model_performance}
% \end{table}

\subsubsection{RNA-RNA-Binding Protein interaction prediction}

\textbf{CLIP database}. The CLIP database identifies experimentally validated protein-RNA interactions from several experimental sources \citep{CLIPdb,eclip,postar3}. Here, we used it to generate a dataset of pairs of genomic-context RNA sequences that share a binding site for either identical or different RNA-binding proteins (RBPs), forcing the predictive model to look for possibly degenerate motifs that repeat in both sequences. To keep the difficulty of the task manageable, we selected five proteins with distinct and sufficiently different position weight matrices and abundant hits: CSTF2T, HNRNPM, KHSRP, SF3B1 and U2AF2, and chose the high-resolution eCLIP dataset as a reference. Since between any two protein-specific datasets ca. 2\% of binding sites were found to overlap, we filtered the data to only extract genomic-context windows with a unique RBP binding site. Here, each sequence length was drawn from the uniform distribution between 200 and 250 nucleotides.

\begin{figure}[h!]
    \centering
    \includegraphics[width=\linewidth]{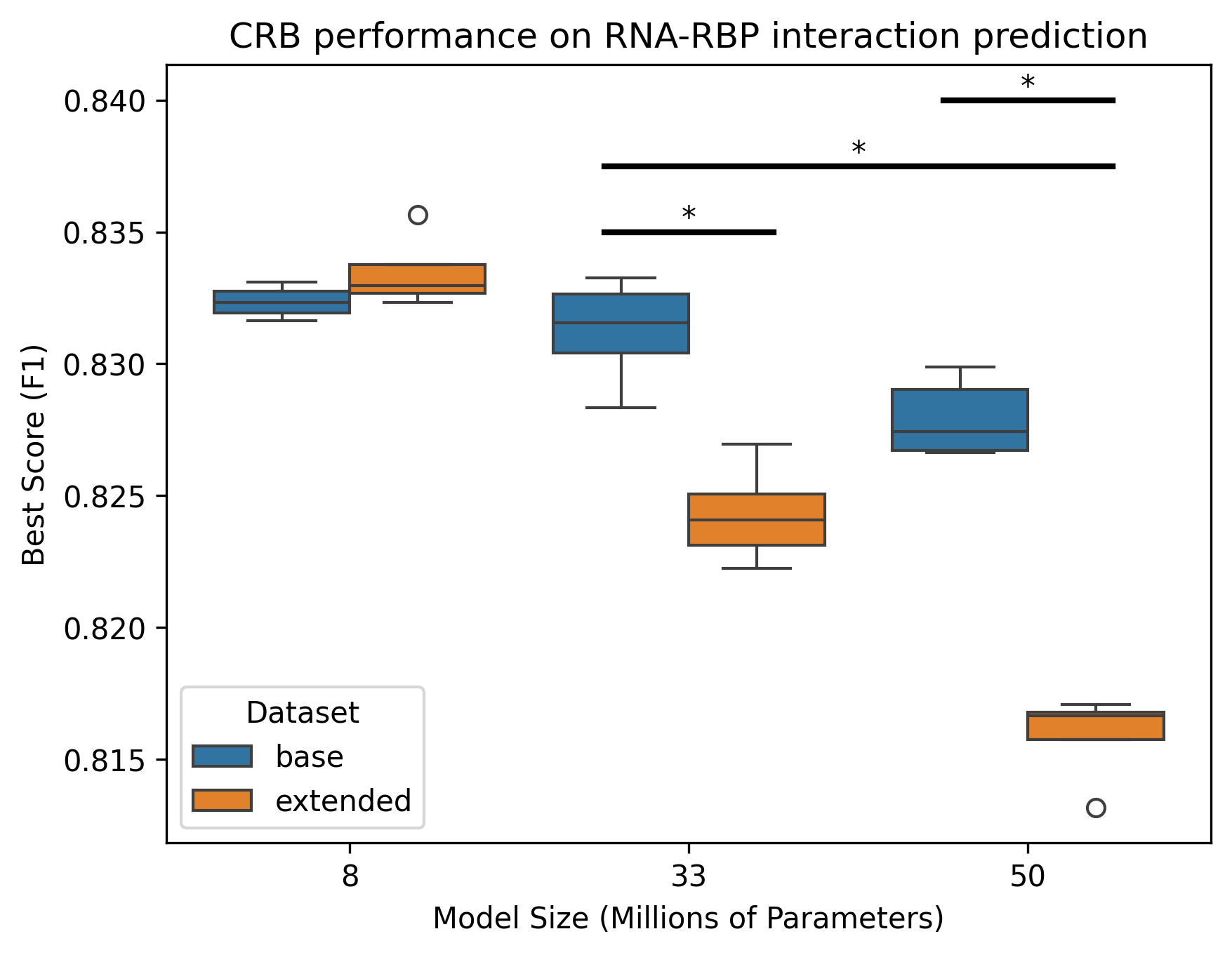}
    \vskip 0.1in
    \caption{Performance of different CRB configurations on the RNA-RBP interaction task.}
    \label{fig:clip_performance}
\end{figure}

We approached this task as a single-label, multi-class classification problem, mapping each sequence to one of five RNA-interacting proteins. The task-specific head consisted of a simple linear layer that mapped the class token to a tensor of size five. All CRB configurations achieved F1 scores above 80, highlighting the model's effectiveness in capturing sequence patterns associated with recognition by these proteins. Figure \ref{fig:clip_performance} presents boxplots for various CRB configurations, grouped by model size and dataset type used in pretraining. We observe that there are statistically significant differences between base and extended datasets for the sizes of 33 and 50 million parameters. This shows that the expansion of the pretraining dataset with coding-RNA sequences is detrimental for performance in this downstream task. When comparing model configurations with different sizes, the only statistically significant difference appears between the 33M model pretrained with the base dataset and the 50M model trained with the extended dataset. With the amount of data available, it is not possible to determine how model size affects models pretrained with the same dataset. However, the general tendency indicates that increasing model size is also detrimental to performance in this downstream task.

Table \ref{table:model_performance_clip} shows the performance of CRB along with other reference models in our study. Comparing CRB with other models, we observe that it significantly outperforms LSTM and CNN baselines, highlighting the advantages of a more complex architecture for enhanced representation in this task. CRB also surpasses RiNALMo by a statistically significant margin, demonstrating the benefits of using GBST as the tokenizer, enabling an 8M parameter model to outperform a 650M model.

\begin{table}[h!]
\centering
\begin{tabular}{lcc}
\textbf{Model} & \textbf{Performance} \\ \hline
LSTM      &  0.719  $\pm$ 0.006  \\ 
CNN        & 0.770 $\pm$  0.054 \\ 
RiNALMo        & 0.831 $\pm$ 0.001 \\ 
ChaRNABERT     & \textbf{0.833} $\pm$ \textbf{0.015} \\ \hline
\end{tabular}
\vskip 0.1in
\caption{F1 performance in the CLIP derived dataset. \textbf{Bold} measures indicate the best model for a dataset under a t-test with p-value $<$ 0.05. }%(p-value = 0.0199).}
\label{table:model_performance_clip}
\end{table}

\subsection{Predicting aptamer-protein interactions}

An RNA aptamer is a short, single-stranded nucleic acid that can selectively bind to a specific target molecule, such as proteins, small molecules, or even entire cells. Aptamers are typically identified through a process known as SELEX (Systematic Evolution of Ligands by EXponential enrichment) \citep{selex}, where a large, randomized library of sequences is screened to find those that bind strongly and specifically to the desired target.
Aptamers are versatile molecules that can bind to various targets, including ions, small organic compounds, proteins, and even cells and viruses. Their specificity makes them highly useful across multiple fields. In diagnostics, they enable the detection of disease biomarkers, such as those in cancer and HIV,  via biosensors. In therapeutics, aptamers are being developed as targeted treatments, like Macugen, approved for age related macular degeneration \citep{ng2006pegaptanib}. As research tools, they help explore molecular interactions due to their precise binding capabilities. Additionally, aptamers can facilitate drug delivery by targeting and delivering therapeutic agents directly to diseased cells \citep{aptamerreview}.

\textbf{University of Texas Aptamer Database.} We leverage the University of Texas Aptamer Database (TAD) \citep{askari2024utexas} to curate a dataset of 2310 pairs of aptamer-protein interactions. The database comprises 1443 aptamer sequences and 561 different protein sequences. Since TAD only encompasses positive examples, we generated negative samples dataset by leveraging sequence dissimilarity as the primary criterion. Specifically, we ensured that the protein sequences selected for negative pairing did not share more than 50\% sequence similarity with any of the positive interaction partners. We then split in training and testing sets by ensuring a maximum of 50\% identity in sequence similarity between training and testing aptamer sequences. %We then split in 2 different training and testing sets by ensuring a maximum of 50\% identity in sequence similarity and secondary structure similarity between training and testing aptamers. 

\begin{figure}[h!]
    \centering
    \includegraphics[width=\linewidth]{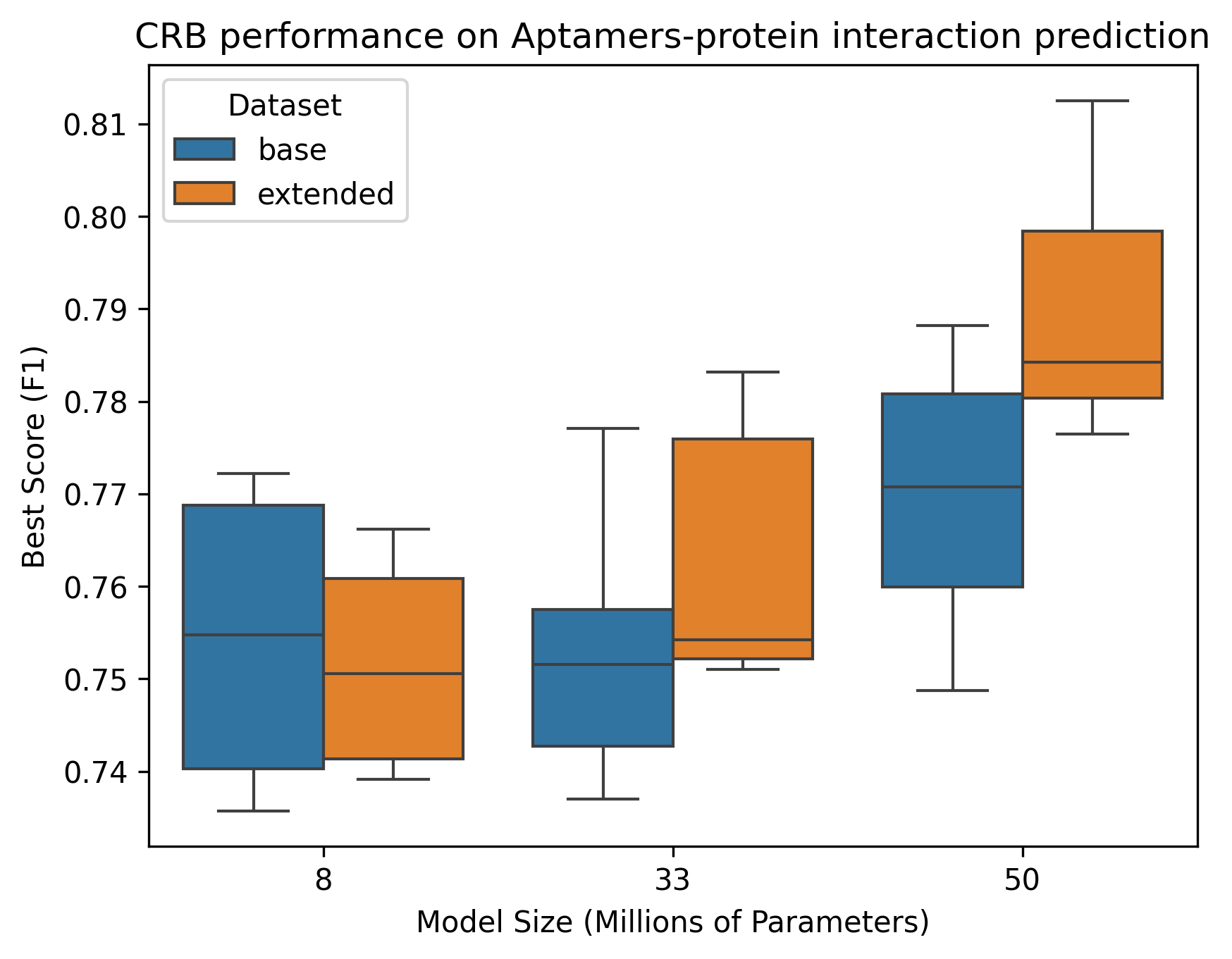}
    \vskip 0.1in
    \caption{Performance of different CRB configurations on the aptamer-protein interaction task.}
    \label{fig:aptamers_performance}
\end{figure}

\begin{table}[h!]
\centering
\begin{tabular}{lcc}
\textbf{Model} & \textbf{Sequence 0.5}  \\ \hline
LSTM        & 0.724 $\pm$ 0.014  \\ 
CNN        &  0.647 $\pm$ 0.024  \\ 
RiNALMo        & 0.744 $\pm$ 0.015          \\ 
ChaRNABERT     & \textbf{0.791} $\pm$ \textbf{0.018}             \\ \hline
\end{tabular}
\vskip 0.1in
\caption{F1 performance in the TAD datasets. \textit{Sequence 0.5} corresponds to model performance in the TAD test set by sequence identity split. \textbf{Bold} measures indicate best model for a dateset under a t-test with p-value $<$ 0.05.}
\label{table:model_performance_apt}
\end{table}

We extracted protein sequence representations using ESM-650M which we used statically. In order to compress the information of the whole representation into a single dimensional tensor we used 10 blocks of residual convolution followed by an adaptive average pooling. The aptamer sequence was processed by CRB and the class token was used for the following steps. Finally, the protein and RNA derived tensors are concatenated and passed through a linear layer. We used a binary cross entropy as the loss function and the F1 as main performance metric following the choices performed in the BEACON benchmark. 

Figure \ref{fig:aptamers_performance} shows the performance of different CRB configurations grouped by pretraining dataset and model size. Despite no statistically significant difference is found between any of the groups, a trend is observed. Model performance tends to increase with model size, suggesting that for this task in particular may be beneficial to further increase the size of the model. Interestingly, the CRB architecture with 50 million parameters trained with extended pretraining dataset already surpasses RiNALMo's performance, suggesting that GBST greatly compensates the need for parameters while still allowing model improvement by increasing them (Table \ref{table:model_performance_apt}).

% and \textit{Structure 0.5} refers to performance by secondary structure similarity split.

%\subsection{RNA-Protein interaction prediction via biomolecular embeddings}

%\subsection{RNA inverse folding}

%\subsection{Prediction of RNA tertiary structure}

%Understanding RNA structure is essential for unraveling the mechanistic basis of its biological functions. Yet, due to its more flexible nature and weaker interactions, very few RNA's have been crystallized and predicting RNA's tertiary structures is still an open problem. \citep{drfold} 
%\subsection{RNA sequence generation}
%RNA therapeutics represent a transformative approach in molecular medicine, offering unprecedented control through programmable nucleic acid sequences allowing us to target proteins, gene and transcripts that were previously thought undruggable. \citep{Rna_therapeutics} 

%This promise of targeting the undruggable has lead to a spike of interest in utilizing generative models for generating novel candidates. Although our model is an encoder-only model and does not natively support generation, we decided to couple it with a latent generation approach to circumvent this limitation. \citep{rna_gen}

%This approach is a two step one, first the embeddings are compressed using a Query Transformer and a sequence decoder is trained to reconstruct RNA sequences from this latent space.  Afterwards a diffusion model is trained to generate latent variables that are then decoded in candidate sequences. Further details can be found at the original material.

\section{Conclusion}

In this study, we introduced ChaRNABERT, a suite of RNA foundational models that employ a learnable character-level tokenization as an inductive bias. Our evaluation across the BEACON benchmark highlights that CRB models achieve competitive or superior performance relative to existing models, often with significantly fewer parameters. 

CRB models performed well in structural prediction tasks, specifically, the CRB-50M model, surpassed larger and specialized models, demonstrating its ability to effectively learn complex structural features directly from sequence data. In functional prediction tasks, CRB models excelled at identifying modeling alternative polyadenylation site usage, essential for insights in post-transcriptional modifications. CRB models demonstrated robust performance across a wide range of tasks, with only a few areas where other models held a slight advantage. For instance, in secondary structure prediction, RiNALMo’s larger parameter count and task-specific training provided a modest advantage. However, these differences were minimal, particularly given CRB’s generalistic training approach and parameter count. %This balance of strong and diversified performance and computational efficiency affirms CRB as a versatile and performant language model for RNA tasks.

In general, increasing the CRB model size parameters did not lead to significant gains in performance across most task categories. This suggests that the advantages of scaling model size may plateau beyond a certain point, allowing CRB models to maintain efficiency without sacrificing accuracy. However, tasks such as non-coding RNA classification and distance prediction did show some performance benefits with larger model sizes, indicating that certain tasks may benefit from targeted scaling. Examining the impact of training on datasets that include both coding and non-coding sequences, we found that combining these sources can enhance CRB's performance, particularly when paired with parameter scaling.

With the aim of expanding current benchmarks, we also devised and implemented two additional RNA tasks: RNA-RBP and aptamer-protein interaction prediction. The RNA-RBP interaction task demonstrated CRB’s superior ability to capture RNA-binding protein motifs and structural features, surpassing both baseline models and RiNALMo in classifying RNA-protein interactions accurately. Similarly, in the aptamer-protein interaction prediction task, which involved synthetic RNA molecules engineered for high-affinity binding, CRB outperformed both standards. These results confirm that CRB models are versatile foundational models, capable of handling a broad range of RNA-related tasks across both natural biological processes and synthetic biology applications, including drug discovery.

CRB models were able to achieve a strong and consistent performance across various tasks, with only slight differences observed in specific areas where models, with larger parameter counts and task-specific training, had a modest edge. Our findings emphasize that rethinking tokenization strategies is crucial when modeling biomolecules, as similar sequence patterns can mask vastly different functions and behaviors. The use of GBST provides a powerful inductive bias, enabling a straightforward BERT architecture to achieve competitive or superior performance with a fraction of the parameters required by other models. We hope this work sets a new direction for RNA language models, with a stronger focus on adaptable sequence representation that minimizes reliance on manually defined tokenization choices.

\section{Future Work}

This preprint is a first version of this work. We aim to expand this manuscript with downstream tasks related to general RNA-protein interaction prediction, RNA inverse folding, RNA folding, and RNA sequence generation. We will also provide a more detailed comparison of model sizes, specifically 150M and 650M and the differences in performance of MLM and UL2.

\section*{Acknowledgments}

We extend our gratitude to the PRACE-EuroHPC Joint Undertaking initiative for establishing competitive compute calls accessible to European SMEs. Part of this work was conducted under an AI and Data Intensive Applications call, awarded through the CINECA HPC network. Additionally, we would like to thank the Barcelona Supercomputing Center for organizing a competitive call to provide Spanish SMEs with access to MareNostrum 5 resources, which have also supported parts of this work. Thanks to S.T. for her support with graphics.

\bibliography{example_paper}
\bibliographystyle{icml2024}

%%%%%%%%%%%%%%%%%%%%%%%%%%%%%%%%%%%%%%%%%%%%%%%%%%%%%%%%%%%%%%%%%%%%%%%%%%%%%%%
%%%%%%%%%%%%%%%%%%%%%%%%%%%%%%%%%%%%%%%%%%%%%%%%%%%%%%%%%%%%%%%%%%%%%%%%%%%%%%%
% APPENDIX
%%%%%%%%%%%%%%%%%%%%%%%%%%%%%%%%%%%%%%%%%%%%%%%%%%%%%%%%%%%%%%%%%%%%%%%%%%%%%%%
%%%%%%%%%%%%%%%%%%%%%%%%%%%%%%%%%%%%%%%%%%%%%%%%%%%%%%%%%%%%%%%%%%%%%%%%%%%%%%%
\newpage
\appendix
\onecolumn

\section{BEACON tasks implementations} \label{beacon_appendix}

In the implementation of downstream tasks of the BEACON RNA benchmark, we followed the established methodology available in the benchmark’s GitHub repository. All tasks utilized the same dataset, except for the splice site prediction task. For this particular task, we aligned with the dataset and task objective outlined in RiNALMo to enable direct comparison with a potentially stronger baseline. For all other tasks, we applied preprocessing protocols identical to those in the BEACON implementation, with one exception: sequence encoding was consistent with the encoding method used in pretraining.

% TODO: This approach may be revised when secondary structure information is incorporated.
For task-specific modules, we used the same prediction heads as BEACON for most tasks. However, for structure-related tasks such as contact and distance map prediction, we adopted the head developed in RiNALMo for secondary structure prediction. This choice allows for comparison with a potentially stronger reference model, leveraging a task-specific module that has been optimized and validated for structural prediction. We kept these task-specific modules minimalistic to reduce the potential impact of module complexity, thus emphasizing the quality of the foundational model's sequence embeddings.

In all tasks, we employed the evaluation metric used in the main reference model. Thus, for most cases, we relied on the BEACON benchmark’s metrics, with the exception of the splice site prediction task. For this task, we used the F1 score from RiNALMo’s framework instead of top-k precision to align with the framing used in RiNALMo.

% TODO: This approach may be updated once secondary structure prediction modules are included.
To ensure training convergence, we implemented an early stopping protocol. We monitored a smoothed version of the validation loss with a patience threshold of 50,000 steps. The smoothed metric was achieved by applying an exponential moving average with an alpha of 0.1. This patience threshold was selected based on the distribution of steps during which the maximum validation score was maintained before being surpassed, thereby ensuring that training reached the absolute best score by preventing premature stopping.
% TODO: The following approach will be reviewed once secondary structure modules are implemented.
In some downstream tasks, there was an observed divergence between the evaluation metric and the validation loss, particularly under overfitting conditions. In these cases, the validation loss would increase while the evaluation metric continued to improve. To address this, we monitored a smoothed version of the evaluation metric for early stopping, ensuring that the recorded performance accurately reflected the model's capabilities for each configuration.

% TODO: This section may change once training for secondary structure modules is implemented.
As for the datasets used, the BEACON benchmark largely shares its datasets with RiNALMo for tasks such as secondary structure prediction and ribosome loading. However, in the splice site prediction task, the dataset and goal differ between the two publications. In RiNALMo, sequences are classified based on the presence or absence of splice sites, whereas in BEACON, each nucleotide in the sequence is classified into three categories: splice site donor, acceptor, or neither. To facilitate a fair comparison, we adopted the RiNALMo dataset and task framing, providing a stronger benchmark for evaluating model performance.

In certain downstream tasks framed as classification problems, class imbalance posed a challenge for effectively learning patterns in the data. In cases where this imbalance hindered performance, we applied a correction to the loss function by weighting sample contributions inversely to their frequency. This adjustment helped the model prioritize positive samples, even when they were underrepresented in the dataset.

\begin{table*}[h!]
\centering
\begin{tabular}{|c|l|l|p{5cm}|}
\hline
\textbf{Abbreviation} & \textbf{Full Task Name} & \textbf{Task Category} & \textbf{Metric (Abbreviation)} \\ \hline
SSP & Secondary Structure Prediction & Structure & F1 Score (F1) \\ \hline
CMP & Contact Map Prediction & Structure & Precision at Length (P@L) \\ \hline
DMP & Distance Map Prediction & Structure & Coefficient of Determination ($R^2$) \\ \hline
SSI & Structural Score Imputation & Structure & Coefficient of Determination ($R^2$) \\ \hline
SPL & Splice Site Prediction & Functional & Accuracy at K (ACC@K) \\ \hline
APA & APA Isoform Prediction & Functional & Coefficient of Determination ($R^2$) \\ \hline
NcRNA & Non-coding RNA Function Classification & Functional & Accuracy (ACC) \\ \hline
MRL & Mean Ribosome Loading & Functional & Coefficient of Determination ($R^2$) \\ \hline
Modif & RNA Modification Prediction & Functional & Area Under Curve (AUC) \\ \hline
VDP & Variant Distance Prediction & Engineering & Mean Columnwise Root Mean Square Error (MCRMSE) \\ \hline
PRS & Prediction of RiboSwitches & Functional & Coefficient of Determination ($R^2$) \\ \hline
CRI-On & CRISPR On-Target Prediction & Engineering & Spearman Correlation (SC) \\ \hline
CRI-Off & CRISPR Off-Target Prediction & Engineering & Spearman Correlation (SC) \\ \hline
\end{tabular}
\vskip 0.1in
\caption{Summary of BEACON Benchmark Tasks, Categories, and Metrics}
\label{tab:beacon_tasks_legend}
\end{table*}

\newpage

\section{FLOPS forward pass computation} \label{flops}
We follow the protocol of \citep{hoffmann2022training} Embedding matrices are counted in both FLOPS and parameter counts, while non-linearities, biases and layer normalizations are omitted. Due to the small differences in both FLOPs and parameter counts of GBST and non-GBST embeddings, they were omitted from the scaling analysis. 

\begin{table}[htbp]
\centering
\begin{tabular}{ll}
\toprule
\textbf{Operation} & \textbf{FLOPs} \\ 
\midrule
\textbf{GBST} & \\
Embedding layer & $2 \times \text{seq\_len} \times \text{vocab\_size} \times \text{d\_model}$\\
GBST Convolutions & $2 \times( \text{max\_blocksize}^2 \times \text{d\_model} + \text{d\_model}^2)$ \\
GBST Scoring & $2 \times \text{seq\_len} \times \text{d\_model}$\\
\midrule
\textbf{Attention Layer} & \\
KQV projections & $2 \times 3 \times \text{seq\_len} \times \text{d\_model} \times (\text{key\_size} \times \text{num\_heads})$ \\ 
Key @ Query & $2 \times \text{seq\_len} \times \text{seq\_len} \times (\text{key\_size} \times \text{num\_heads})$ \\ 
Softmax & $3 \times \text{num\_heads} \times \text{seq\_len} \times \text{seq\_len}$ \\ 
Softmax @ Query reductions & $2 \times \text{seq\_len} \times \text{seq\_len} \times (\text{key\_size} \times \text{num\_heads})$ \\ 
Output projection & $2 \times \text{seq\_len} \times \text{d\_model} \times (\text{key\_size} \times \text{num\_heads})$ \\ 
\midrule
\textbf{FFN Layer} & $2 \times \text{seq\_len} \times (\text{d\_model} \times \text{ffw\_size} + \text{d\_model} \times \text{ffw\_size})$ \\ 
\midrule
\textbf{Total FLOPs} & \text{Embeddings} + \text{num\_layers} $\times$ (\text{Attention Layer} + \text{FFN Layer}) \\ 
\bottomrule
\end{tabular}
\vskip 0.1in

\caption{Forward pass FLOPS computation. The backwards pass is assumed to have twice the amount of FLOPS as the forward pass.}
\end{table}
\newpage
\section{Parameter count}\label{param:comp}
We follow the parameter count schema of \citet{kaplan2020scaling}, removing sub-leading terms such layer normalizations and biases.
\begin{table}[ht]
    \centering

    \begin{tabular}{c|c}
        \toprule
        \textbf{Operation} & \textbf{Parameters} \\
        \midrule
         \textbf{GBST} & \\
Embedding layer & vocab\_size x d\_model \\
GBST Convolutions & d\_model x max\_blocksize + $\text{d\_model}^2$\\
GBST Scoring & d\_model\\ 
        \midrule
        \textbf{Attention Layer} & \\
        
        KQV projections & 3 x d\_model x (key\_size x num\_heads)\\
        Output projection & d\_model x (key\_size x num\_heads)) \\
        \midrule
         \textbf{FFN Layer} & 2 x (d\_model x ffw\_size) \\
         \midrule
         \textbf{Linear Language Head} & d\_model x vocab\_size \\
         \midrule
         \textbf{Total Parameters} & Embeddings + num\_layers x (Attention Layer + FFN Layer) + RoBERTa Head\\
        \bottomrule
    \end{tabular}
    \vskip 0.1in
    \caption{Parameter computation}
    
    \label{tab:fits}
\end{table}

\section{Scaling law from \citep{hoffmann2022training}}\label{scaling}

Following the methodology outlined in \citep{hoffmann2022training}, we estimated the parameters \((A, B, E, \alpha, \beta)\) of the proposed scaling law:

\begin{equation}
    L(N, D) = E + \frac{A}{N^\alpha} + \frac{B}{D^\beta},
\end{equation}

To achieve this, we applied the optimization procedure recommended by \citep{hoffmann2022training}, minimizing the Huber loss \citep{huber1992robust} to account for the difference between the predicted and observed logarithmic losses. This was done using the L-BFGS algorithm \citep{nocedal1980updating}, which is well-suited for optimizing smooth, differentiable functions.

\begin{equation}
    \min_{A, B, E, \alpha, \beta} \sum_{i} \text{Huber}\left(\log L(N_i, D_i) - \log L_{\text{observed}, i}\right),
\end{equation}

To solve this optimization problem, we first conducted a grid search to explore a range of initial parameter values. The results from the grid search were then refined using L-BFGS, which efficiently minimized the function.

For the exponents \(\alpha\) and \(\beta\), which define how the loss scales with model size and the number of tokens, we used the following relationships derived from the scaling law:

\[
    a = \frac{\beta}{\alpha + \beta}, \quad b = \frac{\alpha}{\alpha + \beta},
\]

where \(a\) and \(b\) are constants determined by fitting the parametric model to the observed data.

\begin{figure}[htbp]
    \centering
    \includegraphics[width=0.5\linewidth]{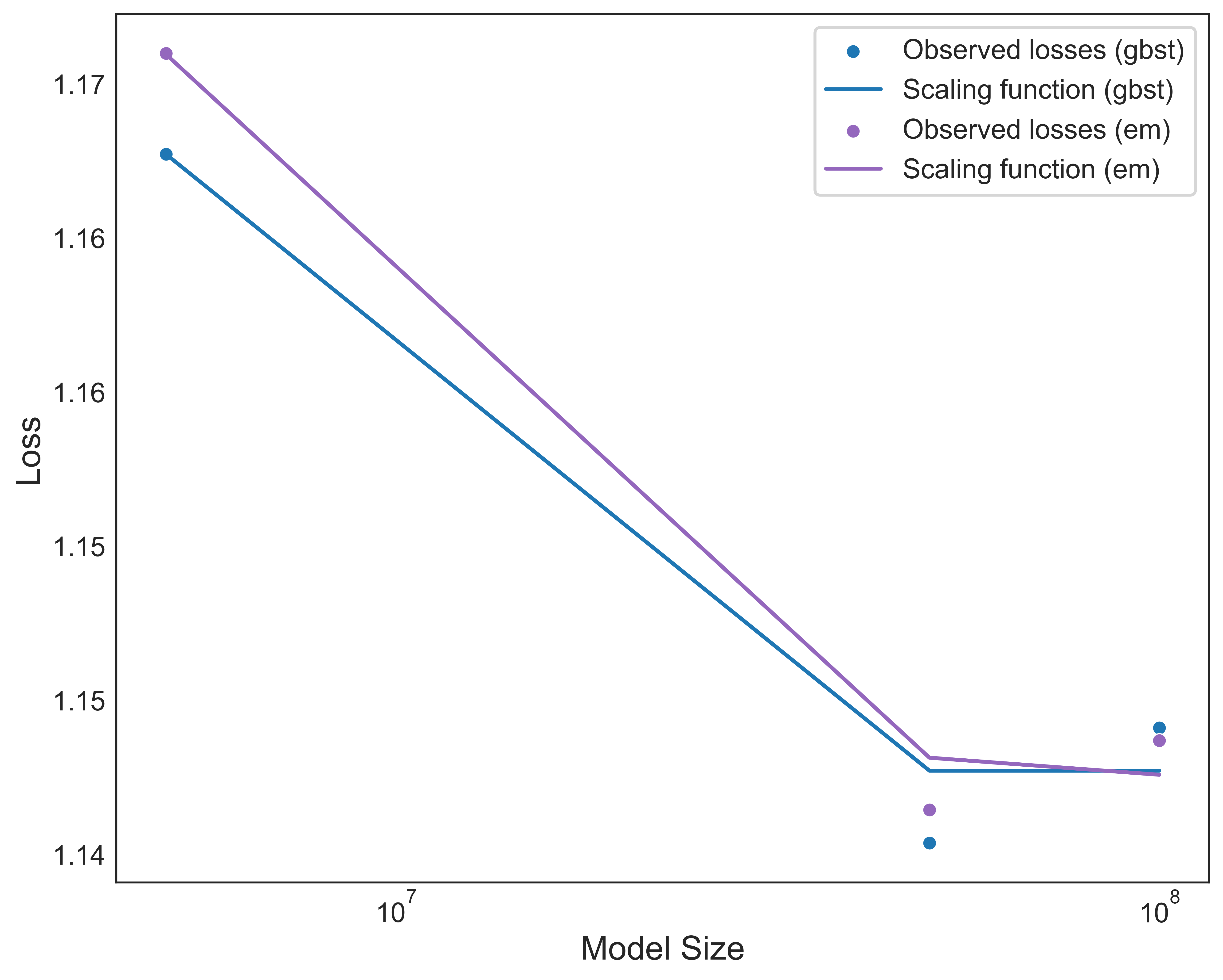}
    \vskip 0.1in
    \caption{Parametric Loss function}
    \label{fig:scaling_law}
\end{figure}

\newpage
\section{CNN, LSTM and RiNALMo configuration and architecture}

We describe the architectures and parameters used in the downstream tasks for the CNN, LSTM and RiNALMo models. All configurations employ the Adam optimizer and follow the strategies described in \ref{extending-bechmark}. Heads used in the downstreams tasks are the same as the ones used in ChaRNABERT.

CNN architecture:

\begin{itemize}
    \item \textbf{Convolutional Layer 1}: 1D convolution with 64 filters, kernel size of 25, padding of 1, and stride of 5. Includes \textit{batch normalization}, followed by ReLU activation and \textit{max pooling} (filter size of 2, stride of 1).
    \item \textbf{Convolutional Layer 2}: 1D convolution with 128 filters, kernel size of 3, padding of 1, and stride of 1. Includes \textit{batch normalization}, followed by ReLU activation and \textit{max pooling} (filter size of 2, stride of 1).
    \item \textbf{Global Average Pooling}: adaptive pooling to reduce the output dimension to 1 per channel.
    \item \textbf{Fully Connected Layer}: 128 neurons with ReLU activation, followed by dropout at a rate of 50\%.
\end{itemize}

\begin{table}[!h]
    \centering
    
    \begin{tabular}{c|ccc}
        \toprule
        \textbf{Parameter} & \textbf{CLIP} & \textbf{Aptamers} \\ \hline
        \textbf{Learning Rate (lr)} & $1 \times 10^{-3}$ & $1 \times 10^{-3}$\\ \hline
        \textbf{Batch Size} & 128 & 128 \\ \bottomrule
    \end{tabular}
    \vskip 0.1in
    \caption{Parameters used for each downstream task in the CNN model}
    \label{table:parameters_cnn}
\end{table}

LSTM architecture:

\begin{itemize}
    \item \textbf{LSTM Layer}: bidirectional LSTM with a hidden size of 128, and 2 layers. Configured with batch\_first as True.
    \item \textbf{Dropout Layer}: with a rate of 50\% 
    \item \textbf{Layer Normalization}: with an output dimension of 128 * 2.
    \item \textbf{Fully Connected Layer}: reduces the output to 128 neurons with a linear transformation.
\end{itemize}

\begin{table}[ht]
    \centering
    \begin{tabular}{c|cc}
        \toprule
        \textbf{Parameter} & \textbf{CLIP} & \textbf{Aptamers} \\ \hline
        \textbf{Learning Rate (lr)} & $1 \times 10^{-4}$ & $1 \times 10^{-3}$ \\ \hline
        \textbf{Batch Size} & 256 & 64 \\ \bottomrule
    \end{tabular}
    \vskip 0.1in
    \caption{Parameters used for each downstream task in the LSTM model}
    \label{table:parameters_lstm}
\end{table}

\newpage
RiNALMo was used in its pretrained version. For all downstream tasks, except for the ones already done in \citep{rinalmo}, we followed the same strategies applied in ChaRNABERT and described in \ref{extending-bechmark}. 

\begin{table}[!ht]
    \centering

    \begin{subtable}[h!]{\textwidth}
        \centering
        \begin{tabular}{c|ccccc}
            \toprule
            \textbf{Parameter} & \textbf{CLIP} & \textbf{Aptamers} & \textbf{CMP} & \textbf{DMP} & \textbf{CRI-On} \\ \hline
            \textbf{Learning Rate (lr)} & $1 \times 10^{-5}$ & $1 \times 10^{-5}$ & $1 \times 10^{-6}$ & $1 \times 10^{-5}$ & $1 \times 10^{-5}$ \\ \hline
            \textbf{Batch Size} & 16 & 8 & 4 & 4 & 1024 \\ \bottomrule
        \end{tabular}

    \end{subtable}

    \vspace{0.3cm} % Space between subtables

    \begin{subtable}[h!]{\textwidth}
        \centering
        \begin{tabular}{c|ccccccc}
            \toprule
            \textbf{Parameter} & \textbf{CRI-Off} & \textbf{VDP} & \textbf{SSI} & \textbf{APA} & \textbf{Modif} & \textbf{NcRNA} & \textbf{PRS} \\ \hline
            \textbf{Learning Rate (lr)} & $1 \times 10^{-6}$ & $1 \times 10^{-5}$ & $1 \times 10^{-6}$ & $1 \times 10^{-5}$ & $1 \times 10^{-6}$ & $1 \times 10^{-5}$ & $1 \times 10^{-5}$ \\ \hline
            \textbf{Batch Size} & 1024 & 16 & 128 & 128 & 256 & 32 & 128 \\ \bottomrule
        \end{tabular}
     
    \end{subtable}

    \caption{Parameters used for each downstream task in RiNALMo}
    \label{tab:parameters}
    
\end{table}

% Except for the head of the ENCORI task for the CNN and LSTM. To avoid overengineering and maintain the architecture of the CNN and LSTM models, the ENCORI head was designed specifically for these two models. The head consisted of a fully connected network with ReLU activation, followed by a linear layer producing an output of dimensions seq\_len×seq\_len×1. As a final layer, a Sigmoid activation is applied, which restricts the output values to the range [0,1], in accordance with the task requirements.

\end{document}